\documentclass[10pt,twocolumn]{article}
\usepackage[top=1.85cm, bottom=1.85cm, left=1.8cm, right=1.8cm]{geometry}
\usepackage{times}  %
\usepackage[hyphens]{url}
\usepackage{graphicx}  %
\usepackage[keeplastbox]{flushend}
\usepackage[hyphens]{url}  %
\usepackage{graphicx} %
\usepackage{tikzsymbols}
\usepackage[sort&compress,numbers]{natbib}

\usepackage{url}                                %
\usepackage{array,multirow,graphicx,adjustbox}  %
\usepackage{booktabs}                           %
\usepackage[utf8]{inputenc}
\usepackage[ruled,algosection,noend,linesnumbered]{algorithm2e}
\usepackage{float}
\usepackage{paralist}
\usepackage{amsmath}
\usepackage{amsfonts}
\usepackage{xcolor}
\usepackage{csquotes} 
\usepackage{tabularx}
\usepackage{makecell}
\usepackage{pbox}
\usepackage[hang,flushmargin]{footmisc}
\usepackage{flushend}
\usepackage{xspace}
\usepackage{multicol, lipsum}
\usepackage[T1]{fontenc}

\usepackage{xcolor}
\usepackage{booktabs}
\usepackage{graphicx}
\usepackage{paralist}
\usepackage[small,bf]{caption}
\usepackage[tight]{subfigure}
\usepackage[hang,flushmargin]{footmisc}
\renewcommand{\footnotesize}{\fontsize{8}{9}\selectfont}
\usepackage[compact]{titlesec}
\titlespacing*{\section}{0pt}{*3}{3pt}
\titlespacing*{\subsection}{0pt}{*2}{2pt}
\usepackage{xspace}

\makeatletter
\def\url@leostyle{%
  \@ifundefined{selectfont}{\def\UrlFont{}}%
  {\def\UrlFont{}}%
}
\makeatother
\usepackage[hyphenbreaks]{breakurl}

\usepackage[bookmarks=true, bookmarksnumbered=true, colorlinks=true, linkcolor=linkcol, citecolor=citecol, urlcolor=urlcol, hypertexnames=true]{hyperref}

\definecolor{darkgreen}{RGB}{0, 100, 0}
\definecolor{linkcol}{rgb}{0.3,0,0}
\definecolor{citecol}{rgb}{0.3,0,0}
\definecolor{urlcol}{rgb}{0.3,0,0}

\makeatletter
\def\url@leostyle{%
  \@ifundefined{selectfont}{\def\UrlFont{\small}}%
  {\def\UrlFont{}}%
}
\makeatother
\newcommand{\descr}[1]{\medskip\noindent\textbf{#1}}
\newcommand{\dspol}{{{\selectfont /pol/}}\xspace}
\newcommand{\travel}{{{\selectfont /v/travel}}\xspace}
\newcommand{\television}{{{\selectfont /v/television}}\xspace}
\newcommand{\fph}{{{\selectfont /v/fatpeoplehate}}\xspace}
\newcommand{\coon}{{{\selectfont /v/CoonTown}}\xspace}
\newcommand{\nigger}{{{\selectfont /v/Nigger}}\xspace}
\newcommand{\redpill}{{{\selectfont /v/TheRedPill}}\xspace}
\newcommand{\deep}{{{\selectfont /v/DeepFake}}\xspace}

\newcommand{\greatawakening}{{{\selectfont /v/GreatAwakening}}\xspace}

\newcommand{\rcoon}{{{\selectfont /r/CoonTown}}\xspace}
\newcommand{\rfph}{{{\selectfont /r/fatpeoplehate}}\xspace}
\newcommand{\rnigger}{{{\selectfont /r/nigger}}\xspace}
\newcommand{\rdeep}{{{\selectfont /r/DeepFakes}}\xspace}

\newcommand{\askvoat}{{{\selectfont /v/AskVoat}}\xspace}
\newcommand{\funny}{{{\selectfont /v/funny}}\xspace}
\newcommand{\politics}{{{\selectfont /v/politics}}\xspace}
\newcommand{\news}{{{\selectfont /v/news}}\xspace}

\newcommand{\reduce}{\vspace*{-0.05cm}}

\let\OLDthebibliography\thebibliography
\renewcommand\thebibliography[1]{
  \OLDthebibliography{#1}
  \setlength{\parskip}{0pt}
  \setlength{\itemsep}{1pt plus 0.2ex}
}

\newif
\ifcomment
\commenttrue
\ifcomment
\newcommand{\sz}[1]{{\bf \textcolor{brown}{SZ: #1}}}
\newcommand{\edc}[1]{{\bf \textcolor{red}{EDC: #1}}}
\newcommand{\gs}[1]{{\bf \textcolor{green}{GS: #1}}}
\newcommand{\ap}[1]{{\bf \textcolor{blue}{AP: #1}}}
\newcommand{\jbnote}[1]{{\bf \textcolor{magenta}{JB: #1}}}
\else
\newcommand{\sz}[1]{}
\newcommand{\edc}[1]{}
\newcommand{\gs}[1]{}
\newcommand{\ap}[1]{}
\newcommand{\jbnote}[1]{}
\fi

\title{\bf ``Is it a Qoincidence?'': An Exploratory Study of QAnon on Voat\thanks{Published in the Proceedings of 30th The Web Conference (WWW 2021). Please cite the WWW version.}}

\date{\vspace*{-0.2cm}-- iDrama Lab, https://idrama.science --}

\author{
Antonis Papasavva\\[-0.25ex]
\normalsize University College London\\[-0.25ex]
\normalsize antonis.papasavva@ucl.ac.uk
\and
Jeremy Blackburn\\[-0.25ex]
\normalsize Binghamton University\\[-0.25ex]
\normalsize jblackbu@binghamton.edu
\and
Gianluca Stringhini\\[-0.25ex]
\normalsize Boston University\\[-0.25ex]
\normalsize gian@bu.edu
\and
Savvas Zannettou\\[-0.25ex]
\normalsize Max Planck Institute for Informatics\\[-0.25ex]
\normalsize szannett@mpi-inf.mpg.de
\and
Emiliano De Cristofaro\\[-0.25ex]
\normalsize University College London\\[-0.25ex]
\normalsize e.decristofaro@ucl.ac.uk\\
}

\begin{document}

\maketitle

\begin{abstract}
Online fringe communities offer fertile grounds to users seeking and sharing ideas fueling suspicion of mainstream news and conspiracy theories. 
Among these, the QAnon conspiracy theory emerged in 2017 on 4chan, broadly supporting the idea that powerful politicians, aristocrats, and celebrities are closely engaged in a global pedophile ring. 
Simultaneously, governments are thought to be controlled by ``puppet masters,'' as democratically elected officials serve as a fake showroom of democracy.

This paper provides an empirical exploratory analysis of the QAnon community on Voat.co, a Reddit-esque news aggregator, which has captured the interest of the press for its toxicity and for providing a platform to QAnon followers. 
More precisely, we analyze a large dataset from \greatawakening, the most popular QAnon-related subverse (the Voat equivalent of a subreddit), to characterize activity and user engagement. 
To further understand the discourse around QAnon, we study the most popular named entities mentioned in the posts, along with the most prominent topics of discussion, which focus on US politics, Donald Trump, and world events.
We also use word embeddings to identify \emph{narratives} around QAnon-specific keywords. 
Our graph visualization shows that some of the QAnon-related ones are closely related to those from the Pizzagate conspiracy theory and so-called drops by ``Q.'' 
Finally, we analyze content toxicity, finding that discussions on \greatawakening are less toxic than in the broad Voat community.
\end{abstract}

\section{Introduction}
Conspiracy theories typically credit secret organizations or cabals for controversial, world-changing events~\cite{swami2012social}; in many cases, they posit that important political events or economic and social trends are the product of deceptive plots mostly unknown to the general public.
A prominent example relates to the disappearance of Malaysia Airlines Flight MH370, which is alleged to have been taken over by hijackers and flown to Antarctica~\cite{newshub2017aircrash}.

The ability to find like-minded people, at scale, on social media platforms has helped the spread of conspiracy theories, and especially politically oriented ones.
For instance, ``Pizzagate''~\cite{pizzagate} emerged during the 2016 US presidential elections, claiming that Hillary Clinton was involved in a pedophile ring.
Even when widely debunked, conspiracy theories can help motivate detractors and demotivate supporters, thus potentially threatening democracies.

Over the past few years, the ``QAnon'' conspiracy has emerged on the anonymous Politically Incorrect (\dspol) board of 4chan.
In October 2017, a user going by the nickname ``Q'' posted numerous threads claiming to be a US government official with a top-secret Q clearance~\cite{originsqanon}.
They explained that Pizzagate was real and that many celebrities, aristocrats, and elected politicians are involved in this vast, satanic pedophile ring.
Q further claimed that President Donald Trump is actively working against a satanic pedophile cabal within the US government. %
QAnon incorporates many theories together into a broadly defined \emph{super-conspiracy theory}.
QAnon adherents also believe that many world events, including the COVID-19 pandemic, are part of a sinister plan orchestrated by ``puppet masters'' like Bill Gates~\cite{jewsbillgates}.
Zuckerman~\cite{Zuckerman2019QAnonAT} argues that QAnon supporters create a vast amount of material that eventually becomes viral.
E.g., the book ``QAnon: An Invitation to a Great Awakening''~\cite{QAnonbook}, written by QAnon followers, ranked second on the Amazon best-selling books list~\cite{vox2019bestseller}.

After Reddit banned QAnon-related subreddits in September 2018~\cite{wired2018qanon,wp2018banq}, QAnon followers reportedly migrated to Voat.co.
Voat is a news aggregator, structured similarly to Reddit, where users subscribe to different channels of interest known as ``subverses.''
Newcomers are not allowed to create new submissions, but can upvote or downvote submissions and comments, and comment on existing submissions.
Once users manage to get ten upvotes on their comments, they can create new submissions to any subverse.

As with many ``fringe'' platforms (e.g., Gab), Voat was designed and marketed vigorously around unconditional support of freedom of speech against the alleged anti-liberal censorship perpetrated by mainstream platforms.
A year after its creation, HostEurope.de stopped hosting Voat because of the content posted~\cite{cancel_voat_contract} and, shortly after, PayPal froze their account~\cite{paypal}.
In August 2015, Voat was thrust into the spotlight when Reddit banned various hateful subreddits (e.g., \rcoon and \rfph~\cite{wired2015coon,motherboard2015nigger}) and a large number of users %
reportedly migrated over~\cite{newell2016user,verge2015voat,ny2017voat}.
The platform shut down in December 2020, with the owner explaining in a post that he ``cannot keep up.''~\footnote{\url{https://searchvoat.co/v/announcements/4169936}}

\descr{Research Questions.} In this paper, we focus on the QAnon-focused community on Voat. 
More specifically, we set out to answer the following research questions:
\begin{itemize}
\item[\bf RQ1:] How active is the QAnon movement on Voat?
\item[\bf RQ2:] Which words and topics are most prevalent for and best describe the QAnon movement on Voat? 
What narratives are shared and discussed by QAnon adherents? 
\item[\bf RQ3:] How toxic is content posted on QAnon subverses? How does it compare to popular subverses focusing on general discussion?
\end{itemize}

\noindent{\bf Methodology.} To address RQ1, we provide a temporal analysis of the most popular QAnon-focused subverse, \greatawakening, in comparison to a {\em baseline} of four of the most popular subverses (in terms of posting activity) focusing on general discussion: \news, \politics, \funny, and \askvoat.\footnote{As discussed later in Section~\ref{sec:datacollection}, we also identify 16 other subverses related to QAnon but find them to be inactive; thus, we only focus on \greatawakening.}
We also analyze submission engagement and user activity.
Then, we use named entities recognition, topic detection, and word embeddings, along with graph representations of QAnon-specific keywords, to define the narratives around the QAnon movement (RQ2).
Finally, to study toxicity within these communities (RQ3), we use Google's Perspective API~\cite{jigsaw2018perspective} to measure how toxic the posts in our dataset are.

\descr{Main Findings.} Our work provides a first characterization of the QAnon community on Voat, through the lens of \greatawakening. 
This subverse attracts many more daily submissions than the four (popular) baseline subverses.
Indeed, users tend to be quite engaged, with two of the most active QAnon submitters creating over $3.75\%$ of the submissions of the baseline subverses as well. 
Also, we analyze user profile data and find that over $17.6\%$ (2.3K) unique users registered a new account on Voat when Reddit banned QAnon subreddits in September 2018.

Using word embeddings, we visualize words closely related to QAnon-specific keywords. 
The movement still discusses, among others, its predecessor conspiracy theory Pizzagate, the posts by the user Q, and other social media.
The most prominent discussion topics are centered around the US, political matters, and world events, while the most popular named entity of the discussion is Donald Trump.
Finally, we find that the QAnon community on \greatawakening is $16.6\%$ less toxic than on baseline subverses.

\section{Background}\label{sec:background}
In this section, we discuss the history, origins, and beliefs of the QAnon movement.
We also provide a high-level explanation of the main functionalities and features of Voat.

\subsection{QAnon}\label{sec:whatisqanon}

\descr{Origins.} QAnon originates from posts by an anonymous user with the nickname Q.
On October 28, 2017, Q posted a new thread with the title ``Calm before the Storm'' on 4chan \dspol. %
In that thread, and over many subsequent cryptic posts, Q claimed to be a government insider with Q-level security clearance.\footnote{This is the US Dept.~of Energy equivalent to the US Dept.~of Defense top-secret clearance.}
The user declared to have got their hands on documents related to, among other things, the struggle over power involving Donald Trump, Robert Mueller, the so-called ``deep state,'' and the pedophile ring that Hillary Clinton supposedly ran~\cite{guardian2018qanon}.
The deep state is believed to be a secret network of powerful and influential people (including politicians, military officials, and others that have infiltrated governmental entities, intelligence agencies, etc.), that allegedly controls policy and governments around the world behind the scenes, while officials elected via democratic processes are merely puppets.
Q claims to be a combatant in an ongoing war, actively participating in Donald Trump's crusade against the deep state~\cite{See2019FromCT}.  

\descr{Ongoing activities.} 
Q has continued to drop ``breadcrumbs'' on 4chan and 8chan, giving birth to a community named after the anonymous (\emph{anon}) user's nickname, ``QAnon,'' devoted to decoding Q's cryptic messages.
This allows them to figure out the real truth about the evil intentions of the deep state, pedophile rings run by aristocrats, and updates on the war Donald Trump was waging.
Although initially this movement was mostly confined to a small group~\cite{guardian2018qanon}, it has since grown substantially via mainstream social networks like Facebook, Reddit, and Twitter and many QAnon adherents around the world have staged protests~\cite{germanyqanon,ukqanon}.%

\descr{Relevance.} 
Sternisko et al.~\cite{sternisko2020dark} and Schabes~\cite{schabes2020birtherism} argue that conspiracy theories, including QAnon, are extremely dangerous for democracies.
Government officials and media often start or promote such conspiracy theories to benefit their political agendas and interests.
For instance, %
at a Trump 2020 rally, the person that introduced Donald Trump used the QAnon motto ``where we go one, we go all'' to conclude his speech~\cite{sky2020whatisq}.
During the 2020 US Congressional elections (November 3), about 25 US Congressional candidates that somehow expressed their support for the conspiracy appeared on ballots.
From those candidates, two elected US House Representatives publicly endorsed the QAnon movement~\cite{QAnoncongress}.%

Notably, before the 2020 US Presidential Elections, the FBI described the QAnon movement as a domestic terror threat~\cite{sky2020whatisq}, and its followers as ``domestic extremists.''
In fact, on January 6th, a pro-Trump mob stormed the US Capitol claiming that ``Q sent them.''
The insurrection resulted in five deaths~\cite{capitol}.
QAnon followers have been arrested for various crimes, including vandalizing churches as the Catholic church allegedly supports human trafficking, kidnapping children to save them from pedophiles, and attempts to murder Canadian Prime Minister Justin Trudeau~\cite{QAnoncrimes}.
Overall, the history of violence surrounding the movement demonstrates that its radicalized followers pose a real danger.

\descr{QAnon on social networks.} Mainstream social networks like Reddit, Twitter, and Facebook have set to ban QAnon-related groups and conversations.
Reddit banned numerous subreddits devoted to QAnon discussion in 2018~\cite{nbc2018containQ,wp2018banq,stuff2018bansq}, then, Twitter put restrictions on 150K user accounts and suspended over 7K others that promoted this conspiracy theory.
Twitter also reported that they would stop recommending content linked to QAnon~\cite{bbc2020twitterban,verge2020twitterban}.
In October 2020, Facebook banned QAnon conspiracy theory content across all their platforms~\cite{bbc2020bansfb}, with YouTube following shortly thereafter~\cite{youtube2020qanonban}.

\subsection{Voat}\label{sec:whatisvoat}
Voat was a news aggregator launched in April 2014, initially under the name ``WhoaVerse'' and renamed to Voat in December 2014.
As mentioned, the platform shut down on December 25, 2020.

\descr{Main features.} Areas of interest, called ``subverses,'' serve to group posts on Voat.
Similar to Reddit, users can register new subverses on Voat, but this functionality was disabled in June 2020. %
When a user registers a new subverse, they become the \emph{owner} of the subverse.
They can delete it and nominate moderators and co-owners, who can in turn then delete comments and submissions.
Voat limits the number of subverses a user may own or moderate to prevent a single user from gaining outsized influence. 
Newcomers can subscribe to subverses of interest, see, vote, and comment on submissions, but are ineligible to post new submissions at this point.
Voat users refer to themselves as ``goats,'' due to the platform's mascot that resembles an angry goat.

\begin{figure}[t]
    \centering
    \includegraphics[width=0.9\columnwidth]{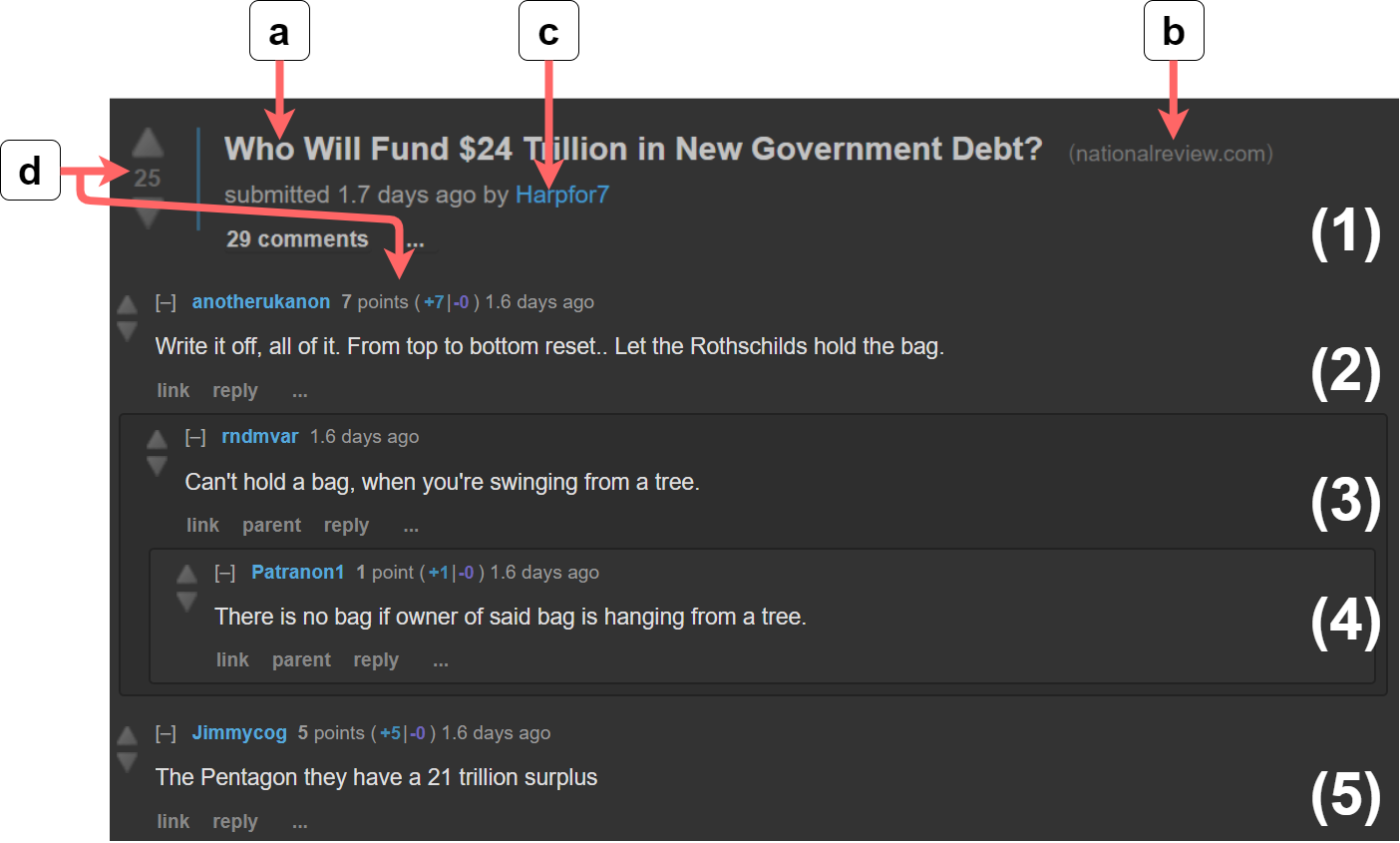}
\reduce\reduce
    \caption{Example of a typical Voat submission. Post with number (1) shows a Voat submission, while posts (2) to (5) are comments.}
    \reduce
    \label{fig:submission-example}
\end{figure}

\descr{Submissions.} Figure~\ref{fig:submission-example} depicts an example of a Voat submission: (1) shows the submission, (2) and (5) are comments made under the submission, and (3) and (4) are child and grandchild of comment (2), respectively. 
A user can create a new submission by posting a title and a description or sharing a link and a description.
If sharing a link, the title of the submission (see ``a'' in Figure~\ref{fig:submission-example}) becomes a hyperlink to the source website.
The source website also appears next to the submission's title (see ``b''), along with the username of the user that posted the submission (``c'').
Note that some subverses allow users to post anonymously.
Other users can then comment on the submission (comment 2 and 5 in Figure~\ref{fig:submission-example}), or comment on comments of other users (comments 3 and 4).
Also, users can ``upvote'' or ``downvote'' the submission or other user's comments (``d'' in the figure).
Submissions and comments may have a negative vote rating based on the votes they receive from users.
A user becomes eligible for posting new submissions only if their \emph{Comment Contribution Points} (CCP) is equal or greater than ten.
The upvotes a user receives are added towards their CCP, while downvotes are subtracted.
Note that users lose their eligibility to post new submissions once their CCP falls under ten.

\descr{Ephemerality.} %
Each subverse has a limit of 500 active submissions at a time: up to 25 submissions in 20 pages (page 0 to page 19).
When a user creates a new submission on Voat, it appears first on page 0, i.e., the subverse's home page.
At the same time, the submission at the end of page 19, usually the one with the least recent comment, is archived.
That submission is still reachable, but only if one knows its direct link; no new comments can be posted to it as it is archived.
When a submission gets a new comment, it is bumped to the top of page 0, no matter when the submission was originally posted, similar to 4chan's ``bumping system''~\cite{papasavva2020raiders}.
However, it is not clear when submissions on Voat stop being bumped when they get new comments.

\section{Data Collection}\label{sec:datacollection}
This section presents our data collection methodology and dataset.

\descr{Subverses.} Our first step is to identify Voat subverses that are related to the QAnon movement. 
To do so, we start from several articles from the popular press~\cite{daily2018deep,wired2015coon,wired2018qanon}, %
which highlight how a few subreddits banned from Reddit re-emerged on Voat.
This happened for QAnon-related subreddits as well~\cite{nbc2018containQ,wp2018banq,stuff2018bansq}; thus, we search for subverses with the same and similar names as the banned subreddits.
We identify 17 subverses and, upon manual inspection, confirm that they are indeed devoted to QAnon-related discussions.
However, we find that 16 out of 17 are essentially inactive, with less than 800 total posts over almost five months.
Therefore, we focus on the most active QAnon subverse, \greatawakening.

We also use the four most active subverses as a baseline dataset.
More precisely, we select the top four, in terms of posts, from the top-10 most subscribed subverses: \news, \politics, \funny, \askvoat.
In the rest of the paper, we refer to these four general-discussion subverses as the ``baseline subverses.''

\descr{Crawling.} We start crawling the five subverses on May 28, 2020, using Voat's JSON API\footnote{\url{https://api.voat.co/swagger/index.html}}, and stop on October 10, 2020.
Voat does not list the archived submissions that fall out of the 20 pages limit, but, as mentioned, these submissions are still reachable if one knows the direct link to it, i.e., the subverse posted in and the submission ID.
A manual inspection of the submission IDs in our database indicates that the submission IDs are monotonically increasing, and thus it is technically possible to collect submissions that fall out of the 20 pages limit by using submission IDs smaller than the ones we collect on the first day that our data collection infrastructure started operating.
If the submission ID does not exist within the subverses we are interested in, the API will return a 404, and thus we could indeed enumerate through all possible submissions.
That said, doing this would require millions of requests to the Voat API, the majority of which would be 404s placing excessive load on their servers, and, if we followed the Voat API usage limits, it would take several years to enumerate through all the possible submissions.

Hence, we use the following methodology to collect all the submissions' comments, focusing only on data posted after May 28, 2020, inclusive.
For each subverse, our crawler continuously requests the submissions pages from 0 to 19. %
We obtain each submission ID, and query the Voat API again to collect the comments posted on that submission.
Voat's API returns only up to 25 comments at a time (aka comment segments) for a given submission.
Next, we note that Voat has a hierarchical, tree-like commenting system, similar to Reddit, with some submissions resulting in branching threads of varying depth.
Thus, to ensure we collect all comments on a submission, our crawler implements a depth first search (DFS) algorithm starting with the comments returned by the first request to the API, and then iteratively query for any child comments they might have.
For each of the children discovered, we query for \emph{their} children until we fully explore the submission's comment tree.
The primary reason we went with a DFS implementation over breadth first search (BFS) implementation is due to the Voat API returning comment \emph{segments}: a DFS simply required a bit less bookkeeping and is a more natural fit considering we are not guaranteed to get all comments at a given level with a single request.
The crawler revisits the pages of every subverse, looking for new submissions, or updates on the ones already collected, numerous times per day, ensuring the collection of the full state of submissions before they fall off the page 19 limit.

\begin{table}[t]
\centering
\small
\smallskip
\begin{tabular}{l r r }
\toprule
\textbf{Subverse} & \textbf{Posts} & \textbf{Users} \\
\midrule
\greatawakening & 152,315 & 4,915\\
\midrule
\news & 153,162 & 6,212\\
\politics & 107,214 & 5,610 \\
\funny & 61,949 & 4,971\\
\askvoat & 35,643 & 4,282\\
\midrule
\textbf{Total} & 510,283 & 13,084\\
\toprule
\end{tabular}
\reduce\reduce
\caption{Number of posts for each subverse in the dataset, along with the total number of user profiles collected.}
\reduce \reduce 
\label{tbl:voat_submissionscomments_crawled}
\end{table}

\descr{Dataset.} Table~\ref{tbl:voat_submissionscomments_crawled} lists the number of posts (submissions and comments) we collect for each subverse analyzed in this study. 
Our dataset spans posts from May 28 to October 10, 2020.
Alas, our dataset is missing some posts between June 9 and June 13 due to failure of our data collection infrastructure.
Besides submissions and comments, we also collect publicly accessible user profile data.
More specifically, we collect profile data of the users posting a submission or a comment on \greatawakening and baseline subverses listed in Table~\ref{tbl:voat_submissionscomments_crawled}.
In total, we find 4.9K, 6.2K, 5.6K, 4.9K, and 4.2K usernames that have either created a submission or made a comment in \greatawakening, \news, \politics, \funny, and \askvoat, respectively.
The union of these results in 15K unique usernames, with 13K of these usernames having accessible profiles.
The remaining $\sim$2K (13.16\%) of usernames we query result in a 404 error, which we believe is due to deleted or deactivated profiles.

\descr{Ethics.} We only collect openly available data and follow standard ethical guidelines~\cite{rivers2014ethical}.
We do not attempt to identify users or link profiles across platforms.
Finally, the collection of data analyzed in this study does not violate Voat API's Terms of Service.

\section{General Characterization}\label{sec:analysis}
This section analyzes aggregate and user-specific activity, content engagement, and registrations for all subverses in our dataset.

\begin{figure}[t!]
\centering
\subfigure[Submissions]{\includegraphics[width=1\columnwidth]{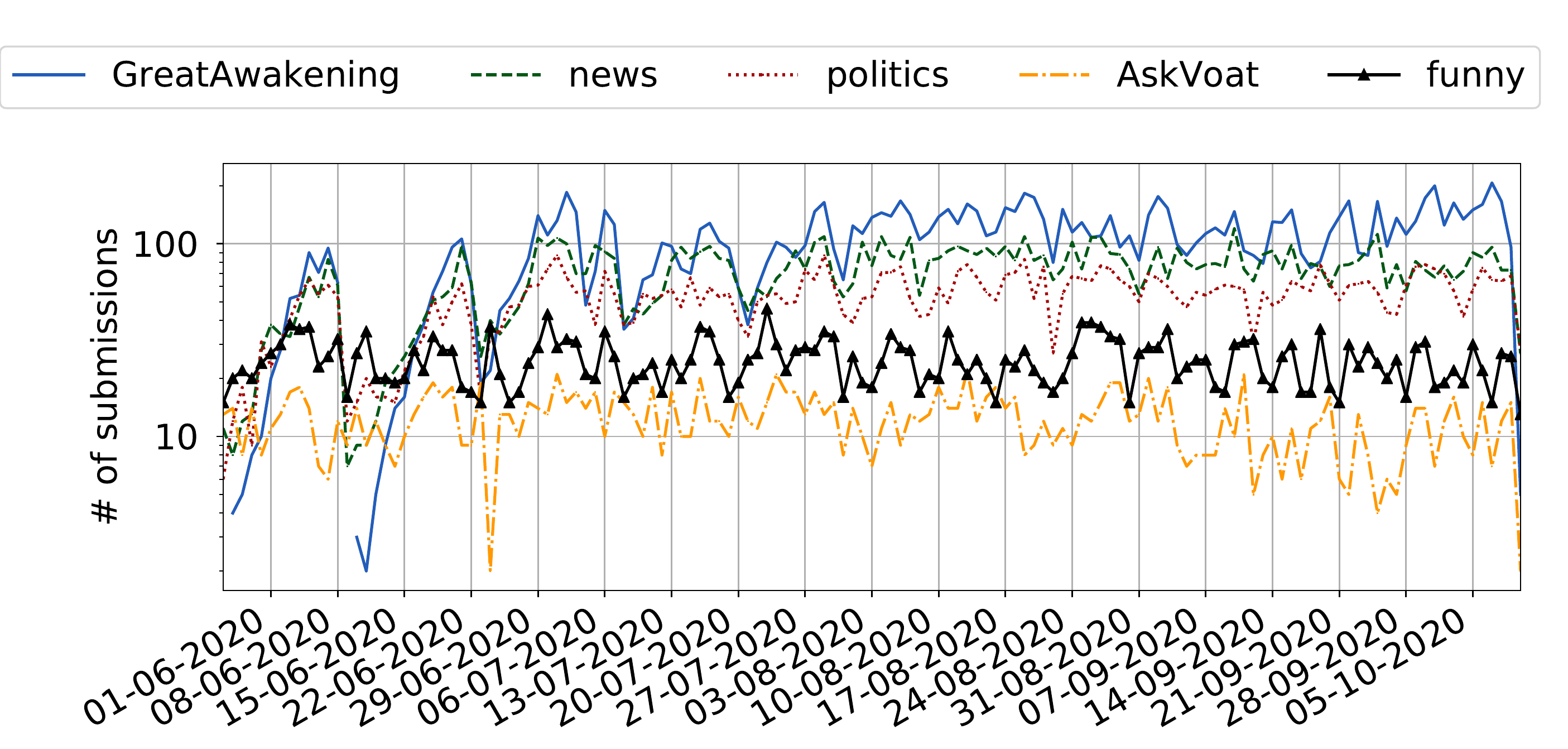}\label{fig:all_submissions_per_day}}\\[-0.5ex]
\subfigure[Comments]{\includegraphics[width=1\columnwidth]{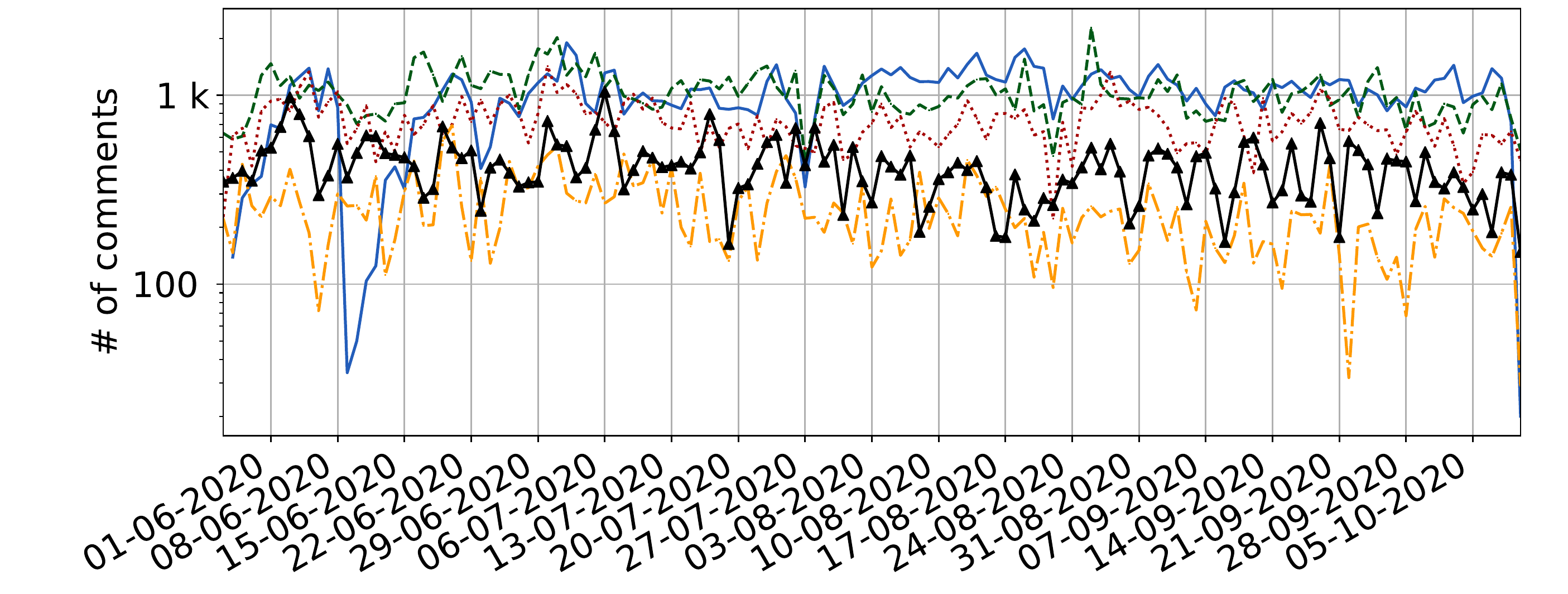}\label{fig:all_comments_per_day}}
\reduce\reduce
\caption{Number of submissions and comments posted per day in baseline subverses and in \greatawakening.}
\label{fig:baselineGA_submissions_comments_per_day}
\reduce
\end{figure}

\begin{figure*}[t!]
\centering
\subfigure[Comments]{\includegraphics[width=0.66\columnwidth]{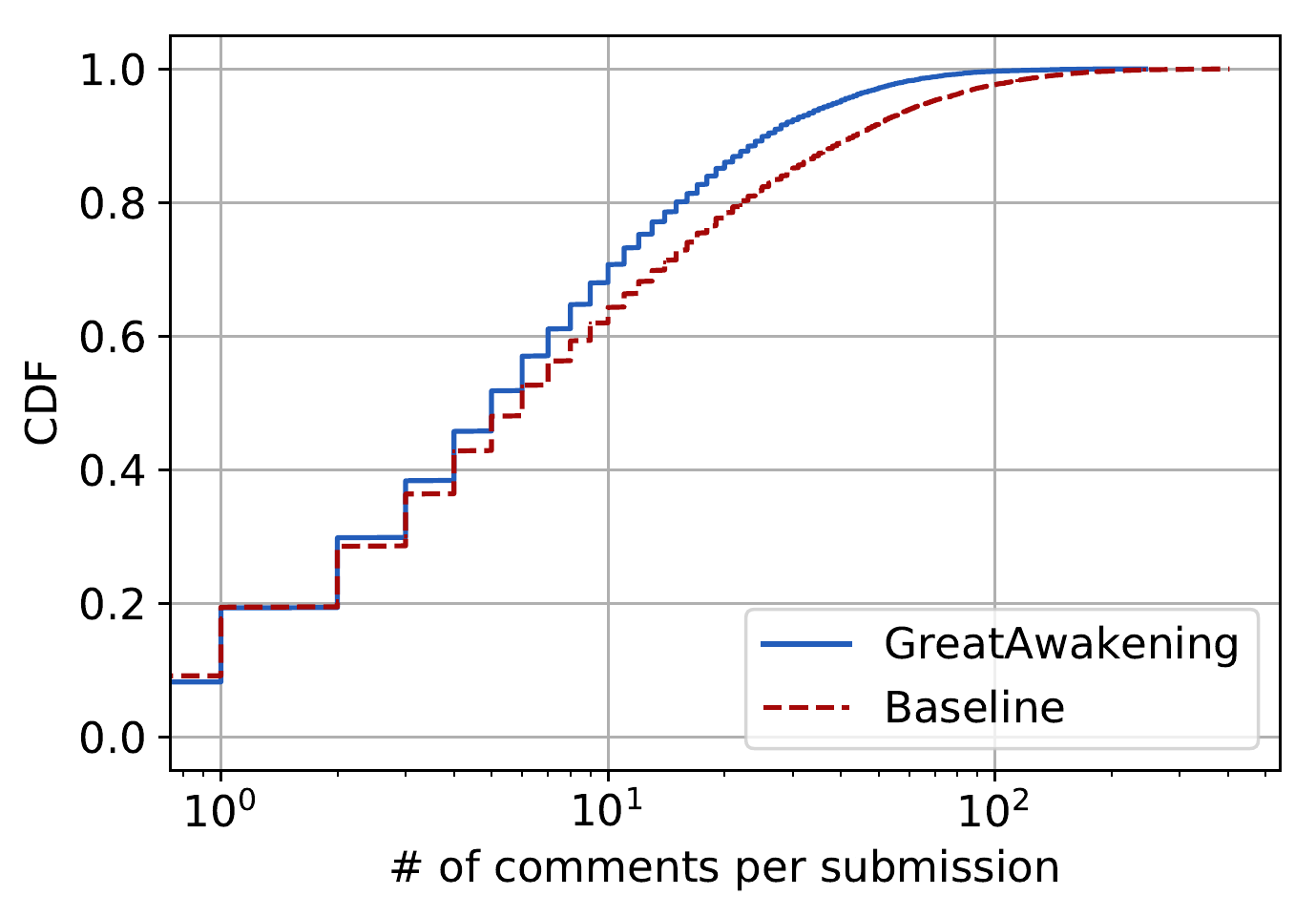}\label{fig:cdf_comments_per_submission}}
\subfigure[Votes]{\includegraphics[width=0.66\columnwidth]{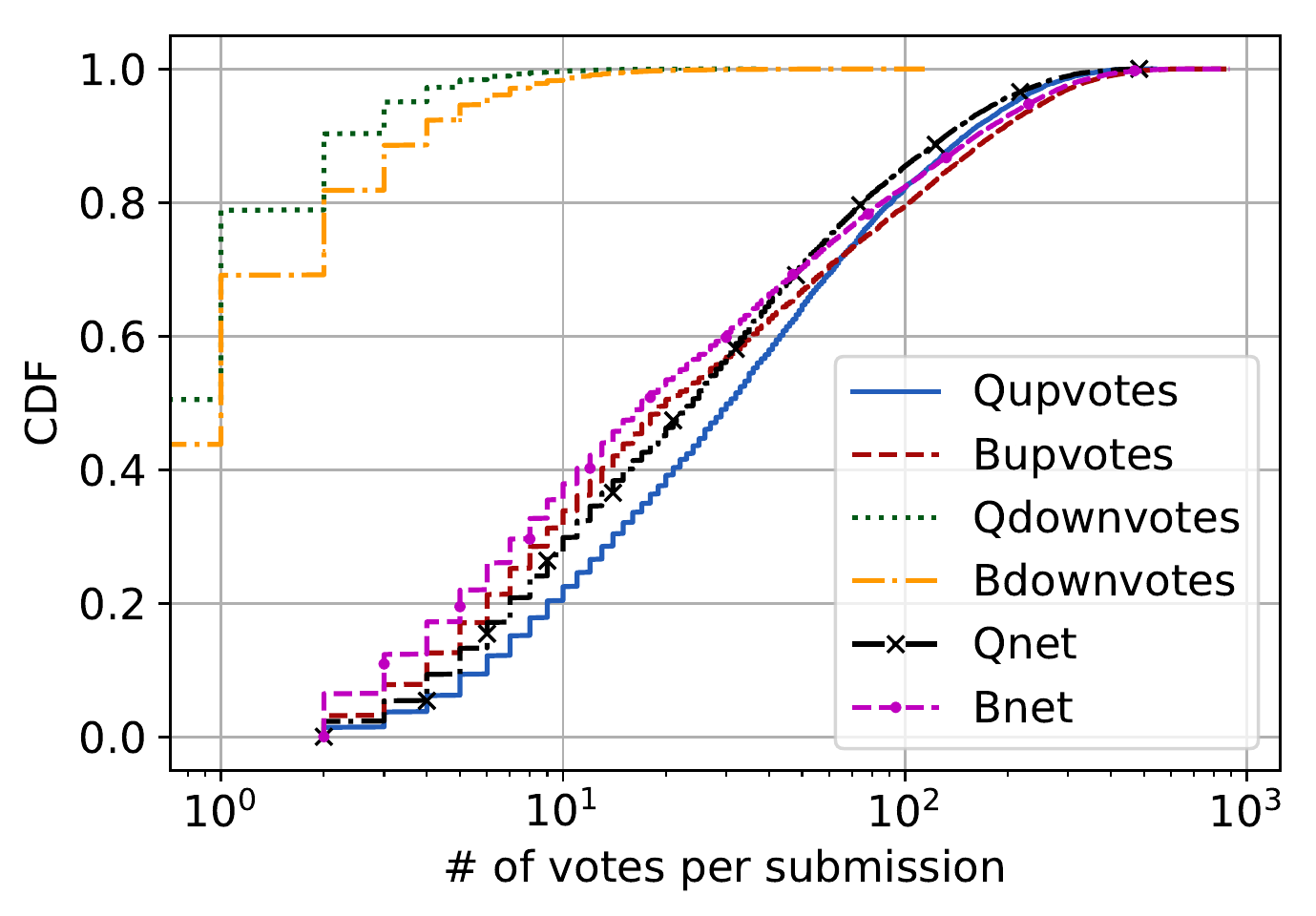}\label{fig:cdf_votes_per_submission}}
\subfigure[Votes]{\includegraphics[width=0.66\columnwidth]{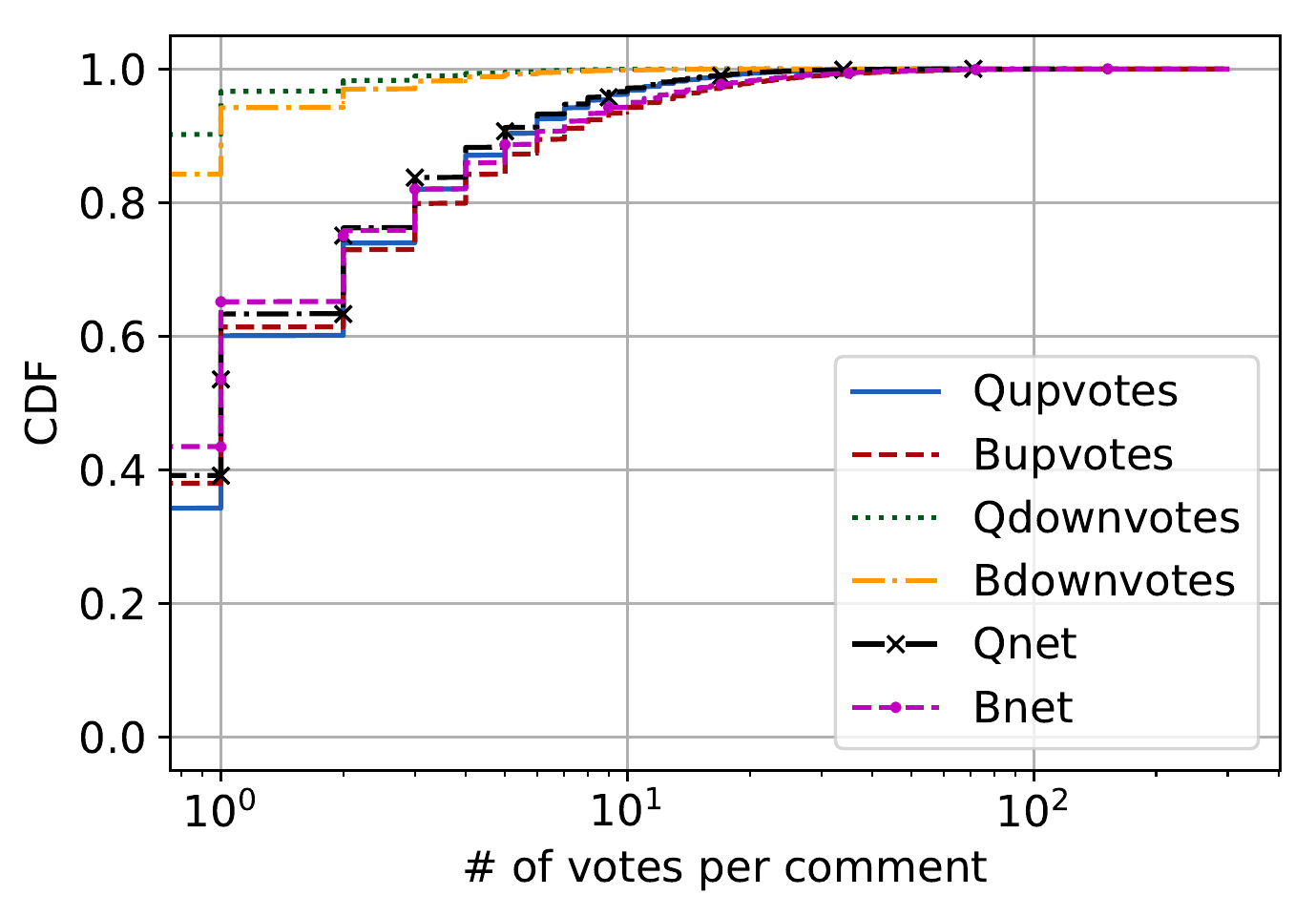}\label{fig:cdf_votes_per_comment}}
\reduce\reduce
\caption{CDF of the number of (a) comments and (b) votes per submission on \greatawakening (Q) and baseline subverses (B), and number of (c) votes per comment on \greatawakening (Q) and baseline subverses (B).}
\reduce
\label{fig:cdf_comments_votes_per_submission}
\end{figure*}

\subsection{Posting Activity}
We start by looking at how often submissions and comments are posted on the collected subverses.
Figure~\ref{fig:all_submissions_per_day} plots the number of daily submissions for the baseline and \greatawakening subverses (note log-scale on the y-axis).
From the figure, we see that over 4.5 months, \greatawakening has more submissions than the individual baseline subverses, with about 100 new submissions per day, on average.
The next most active subverse is \news, with about 70 new submissions per day.
This is remarkable considering that, as of October 2020, \greatawakening has only 20K subscribers, while \news has 100K.
When looking at comment activity (Figure~\ref{fig:all_comments_per_day}), \news and \greatawakening are close, with 1.06K and 1.01K comments per day, respectively.

We observe a peak in submission and comment posting activity on \greatawakening between June 29 and July 3, with the most submissions on July 2 (185 submissions and almost 1.9K comments).
Manual inspection indicates the peak in submission activity may be related to Jeffrey Epstein's ex-girlfriend, Ghislaine Maxwell, being arrested by the FBI~\cite{bbc2020girlfriendarrested}.
Another peak in posting activity appears between August 10 and August 21, with a peak of 183 submissions on August 19. 
Manual inspection does not reveal any apparent link to a specific event.
Finally, October 7 has the most submissions on \greatawakening for a single day (207), which we believe is due to Facebook announcing the ban of QAnon accounts, pages, and related content across all their platforms~\cite{bbc2020bansfb}.

\subsection{Engagement} 
Next, we look at user engagement. %
Figure~\ref{fig:cdf_comments_per_submission} plots the Cumulative Distribution Function (CDF) of the number of comments per submission.
On average, submissions on \greatawakening receive 10.4 comments, while the baseline subverses' submissions get 16.2 comments.
Specifically, Figure~\ref{fig:cdf_comments_per_submission} shows that only $14.9\%$ and $22.3\%$ of the submissions on \greatawakening and baseline subverses, respectively, have more than 20 comments.
The median number of comments on \greatawakening submissions is 5 and on baseline subverse's submissions is 6, while the most popular \greatawakening submission has 245 comments and the most popular baseline subverses' submission has 403 comments.
Our findings show that, although \greatawakening has the most submissions, the users of the baseline subverses are more engaged.

Next, we look at how often users upvote and downvote submissions.
In Figure~\ref{fig:cdf_votes_per_submission}, we plot the CDF of upvotes, downvotes, and net votes (e.g., upvotes - downvotes) the submissions get. 
On average, \greatawakening gets 57.4 upvotes and 0.9 downvotes, while on baseline subverses, we find 61 upvotes and 1.5 downvotes.
The most upvoted submission has 537 and 870 upvotes on \greatawakening and baseline subverses, respectively, while the most disliked submission has 37 downvotes on \greatawakening, and 114 downvotes in the baseline subverses.
Specifically, the title of the most upvoted \greatawakening submission is \emph{``The United States of America will be designating ANTIFA as a Terrorist Organization''} and it links to a tweet by Donald Trump. 
On average, the submissions of both \greatawakening and the baseline subverses tend to have a net positive vote count; about 48.8 for \greatawakening and 54.1 for the baseline subverses.
 
We observe that $62.4\%$ and $50.5\%$ of the \greatawakening and baseline submissions, respectively, have more than 20 upvotes.
On the contrary, only $0.46\%$ and $1.79\%$ of the submissions on \greatawakening and baseline subverses get more than 10 downvotes.
We also run a two-sample Kolmogorov-Smirnov (KS) test on the distributions of upvotes, downvotes, and net votes, and reject the null hypothesis that there is no difference between the distributions ($p < 0.01$ for all comparisons).

Similarly, we plot the CDF of the number of upvotes and downvotes of comments in Figure~\ref{fig:cdf_votes_per_comment}.
On average, comments get 2.2 upvotes and 0.18 downvotes on \greatawakening.
Comments of the baseline subverses get 2.8 upvotes and 0.35 downvotes, on average.
Again, we find statistically significant differences between the distributions via the two-sample KS test ($p < 0.01$).

Overall, this shows that users of both communities tend to vote the content they encounter positively.
Baseline subverses' posts tend to be downvoted and upvoted more often than the \greatawakening posts.
This is probably due to the significant difference in audience between the two communities.
Notably, both communities seem to be engaging towards commenting and voting the posts they encounter on the platform.

\begin{figure*}[t!]
\centering
\subfigure[\greatawakening submissions]{\includegraphics[width=0.2475\textwidth]{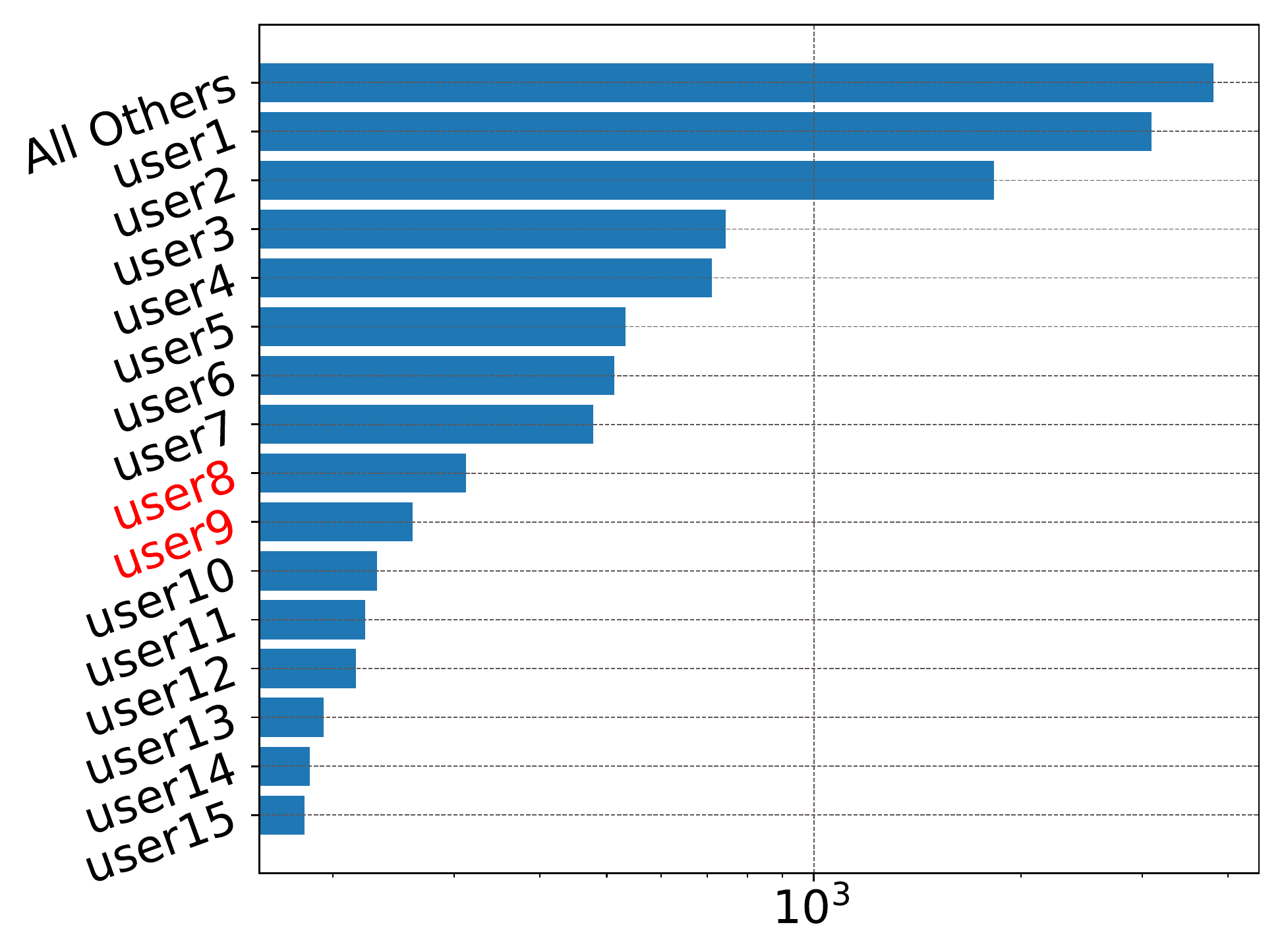}\label{fig:q_submissions_peruser}}\hspace{-0.25cm}
\subfigure[\greatawakening comments]{\includegraphics[width=0.2475\textwidth]{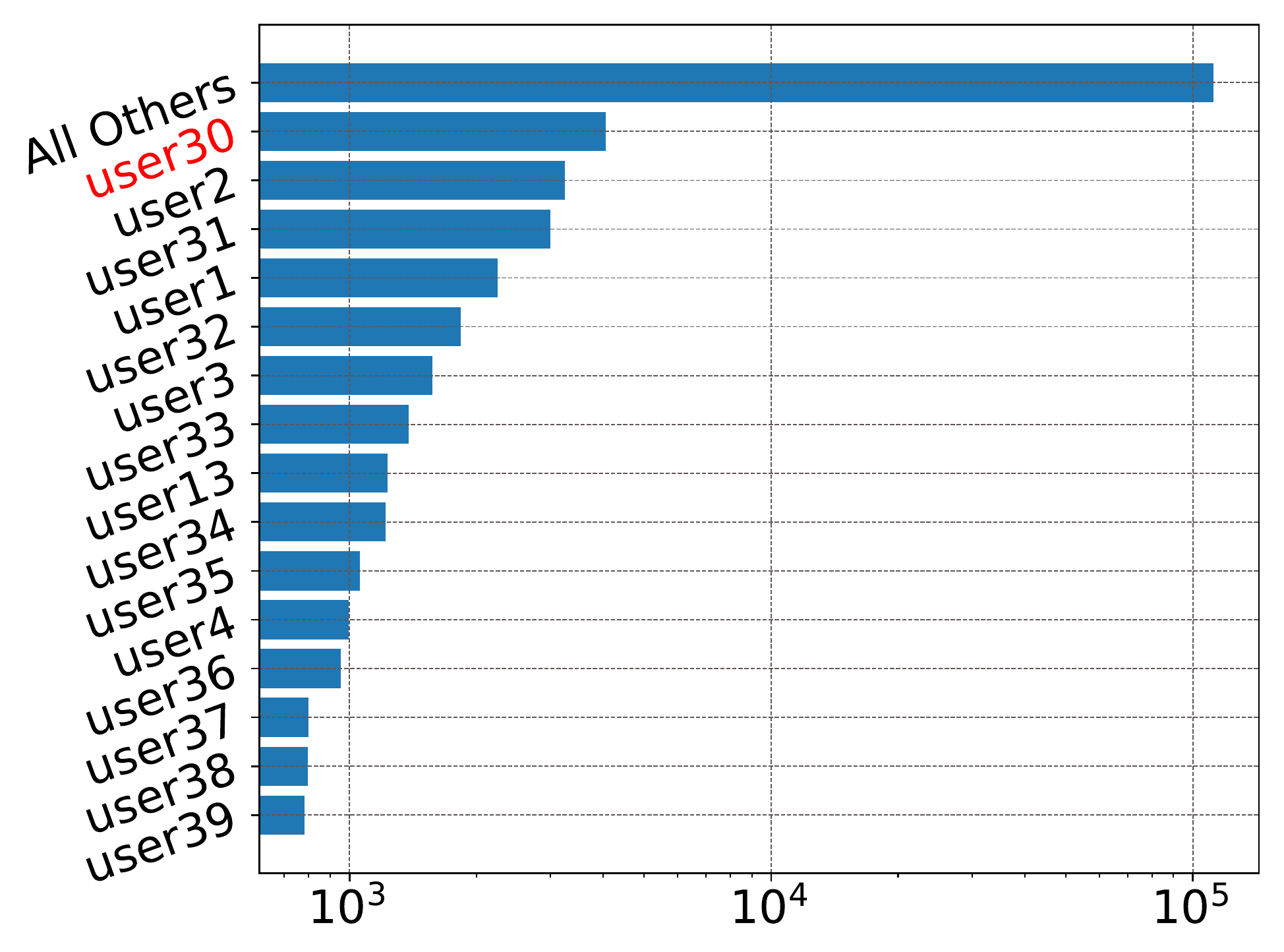}\label{fig:q_comments_peruser}}
\subfigure[Baseline submissions]{\includegraphics[width=0.2475\textwidth]{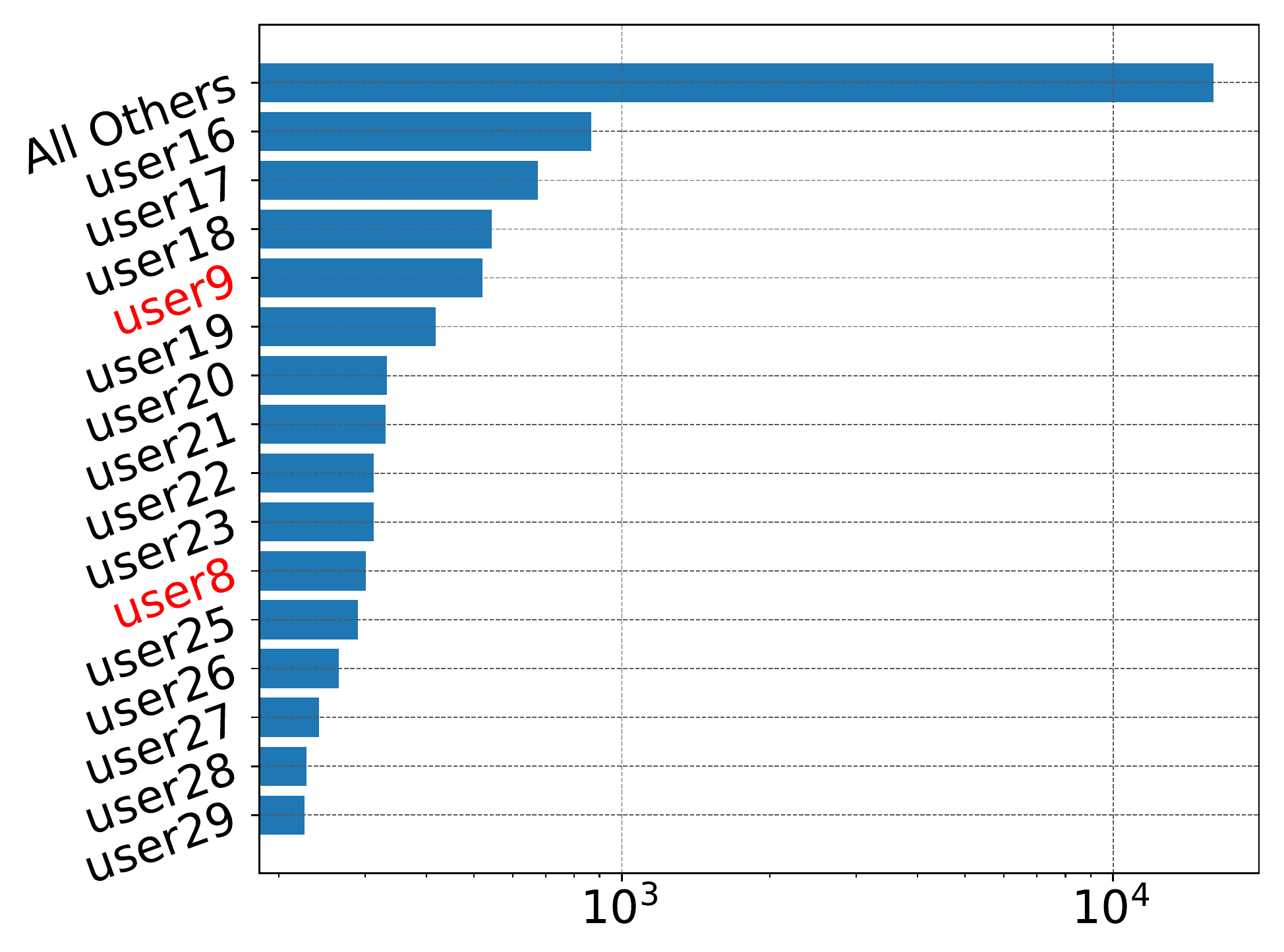}\label{fig:b_submissions_peruser}}
\subfigure[Baseline comments]{\includegraphics[width=0.2475\textwidth]{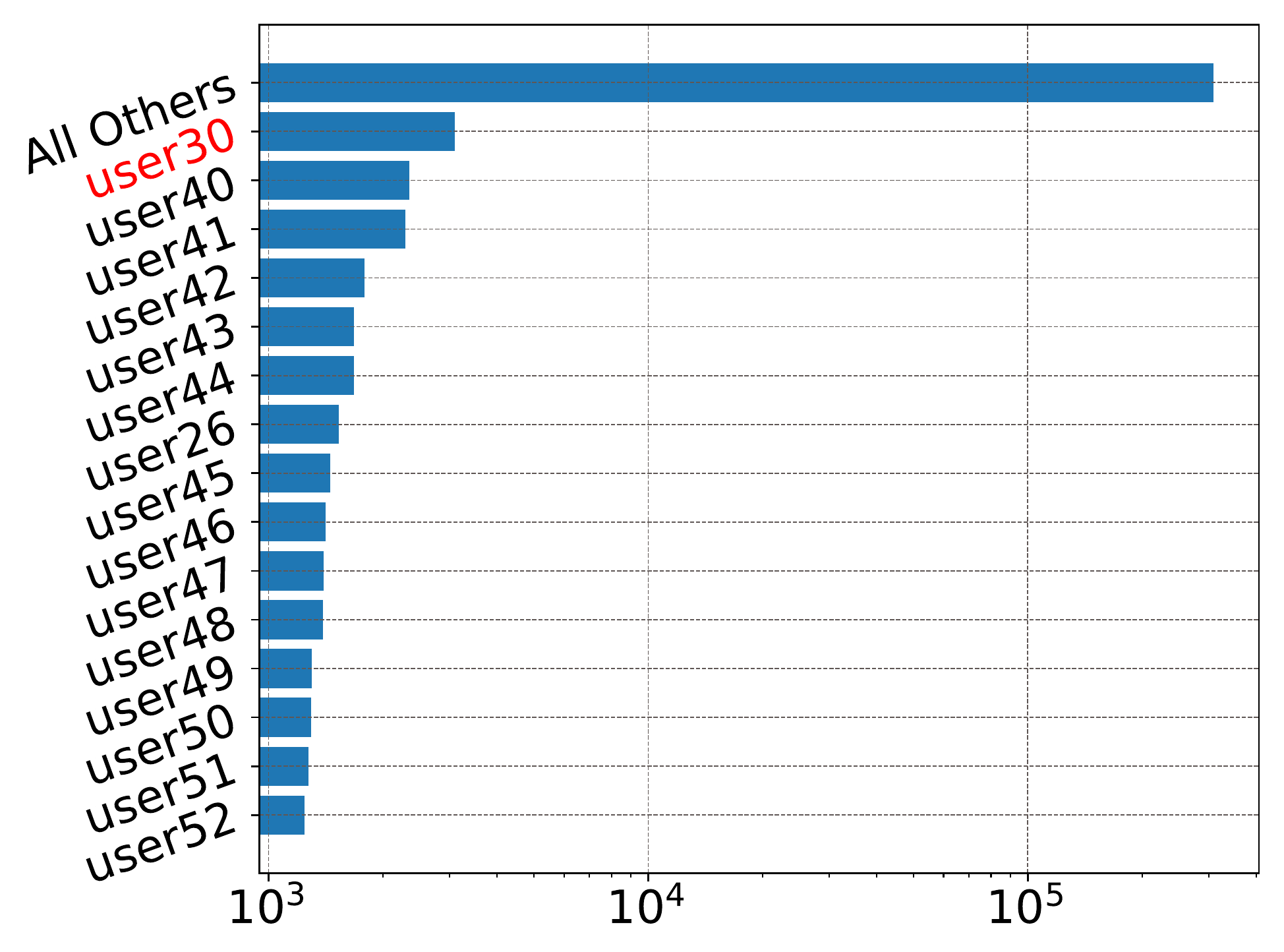}\label{fig:b_comments_peruser}}
\reduce \reduce 
\caption{Number of submissions and comments posted per user on \greatawakening and baseline subverses.} %
\reduce
\label{fig:submissions_comments_peruser}
\end{figure*}

\begin{figure*}[t!]
\centering
\subfigure[\greatawakening user registrations]{\includegraphics[width=1\columnwidth]{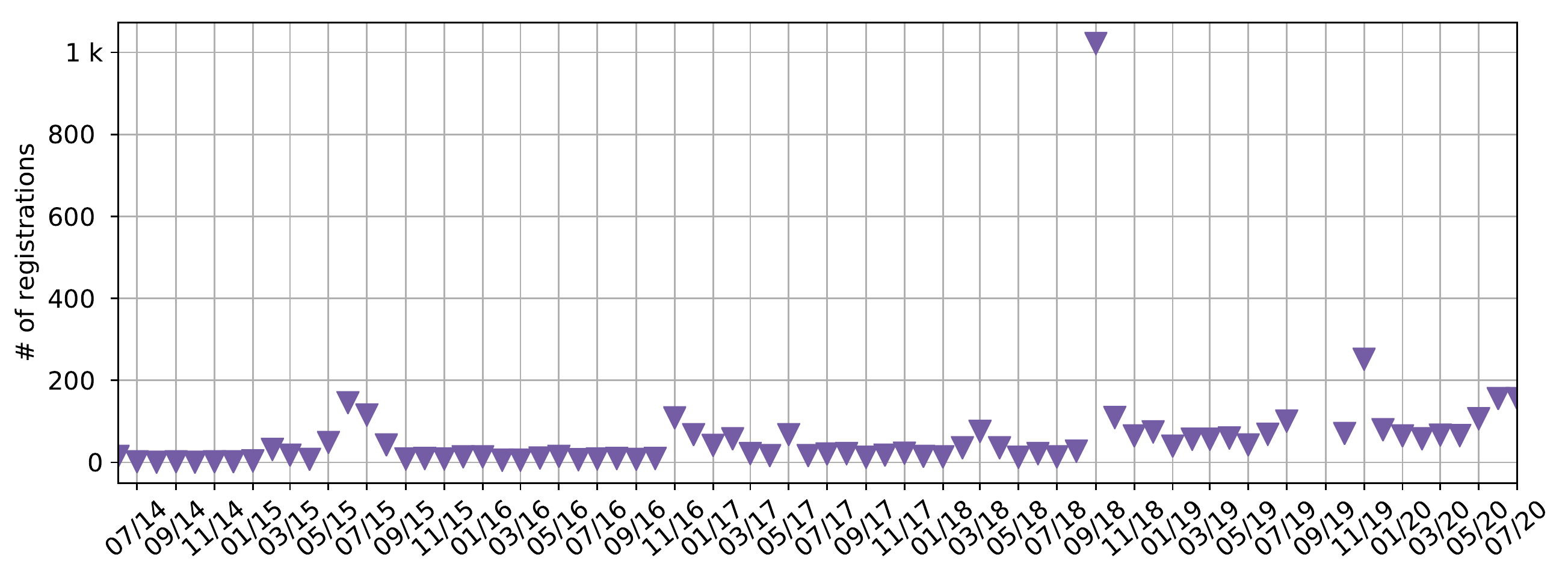}\label{fig:qanon_registrations}} %
\subfigure[Baseline subverses user registrations]{\includegraphics[width=1\columnwidth]{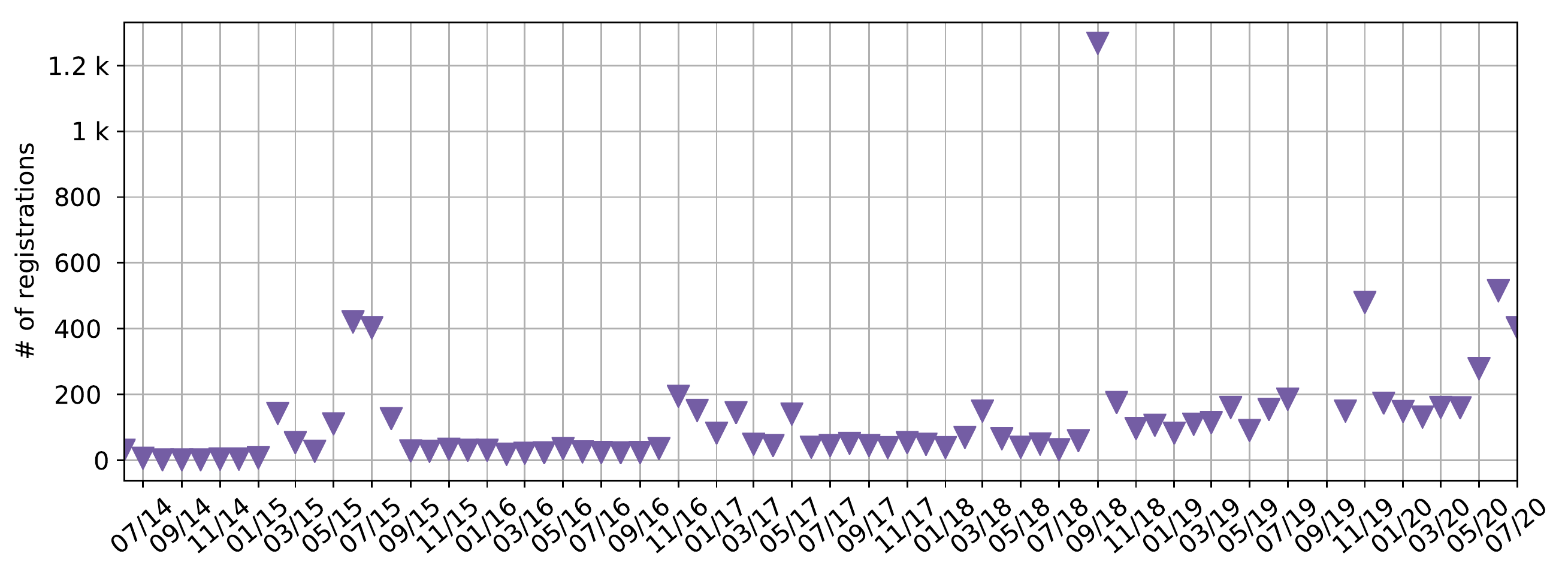}\label{fig:base_registrations}}
\reduce\reduce
\caption{Number of monthly user registrations of (a) users engaging on \greatawakening and (b) users engaging in all baseline subverses.}
\label{fig:registrations}
\reduce
\end{figure*}

\subsection{User Activity}

Next, we focus on user profile data to understand how often users post new submissions.%
More specifically, we investigate whether the audience of \greatawakening and baseline subverses consume information from specific users due to Voat not allowing newcomers posting new submissions unless they achieve a CCP above 10.

To do so, we count the number of submissions users posted on \greatawakening and the baseline subverses.
We find that only 346 users made the 13.5K submissions of \greatawakening.
The 21.9K submissions of the baseline subverses were made by 1.8K users.
Figure~\ref{fig:submissions_comments_peruser} reports the top 15 submitters and commenters of both communities.
To protect users' privacy, we replace the original usernames with ``user1,'' ``user2,'' etc.

We observe that the top submitter, ``user1'' in Figure~\ref{fig:q_submissions_peruser}, posted $22.9\%$ (3.1K) of the total submissions on \greatawakening.
Excluding the top 15 submitters, the remaining 331 submitters (marked as ``All Others'' in the figure) are responsible for $28.2\%$ (3.8K) of the submissions made on \greatawakening.
This is not the case for submissions of general discussion as the top 15 submitters together are only responsible for $26.8\%$ (5.8K) of the total submissions, as depicted in Figure~\ref{fig:b_submissions_peruser}. 
Excluding the top 15 commenters, \greatawakening (Figure~\ref{fig:q_comments_peruser}) and baseline subverses (Figure~\ref{fig:b_comments_peruser}) comment activity seems to fall on the broader audience of the communities since ``All Others'' post $80.9\%$ (112K) and $92\%$ (308.5K) of all the comments, respectively.

Manual inspection of our dataset shows that $22.8\%$ (3K) usernames \emph{overlap} between \greatawakening and the baseline subverses.
Namely, ``user8'' and ``user9'' are amongst the top submitters of both communities, and ``user30'' ranks the first commenter in both.
Our results suggest that the audience of \greatawakening (20K subscribers) consumes content and submissions from a handful of users (349 submitters), and to a great extent, from ``user1.''

\begin{table*}[t]
\centering
\footnotesize
\begin{tabular}{r p{16.5cm}}
\toprule
\textbf{Topic} & {\bf Words per topic for \greatawakening}\\
\midrule
1 & trump (0.004), people (0.003), election (0.003), vote (0.003), child (0.003), pelosi (0.003), biden (0.002), mail\_voting (0.002), voting (0.002), durham (0.002)\\
2 & president\_trump (0.02), president (0.006), trump (0.005), senate (0.005), bill\_gate (0.004), link ( 0.004), gop (0.003), conspiracy\_theory (0.003), video (0.003), bill\_clinton (0.003)\\
3 & new\_york (0.006), deep\_state (0.006), new (0.005), donald\_trump (0.005), debate (0.005), ghislaine\_maxwell (0.004), coronavirus (0.004), trump (0.004), kamala\_harris (0.004), state (0.003)\\
4 & joe\_biden (0.011), supreme\_court (0.007), white\_house (0.007), biden (0.005), potus (0.005), joe (0.005), trump (0.005), post (0.004), fake\_news (0.004), cnn (0.004)\\
5 & trump\_supporter (0.006), kek (0.005), trump\_campaign (0.004), trump (0.004), last\_night (0.004), hunter\_biden (0.004), good (0.003), supporter (0.003), biden (0.003), russia (0.003)\\
6 & united\_state (0.008), mail\_ballot (0.007), nursing\_home (0.006), fox\_news (0.005), social\_medium (0.005), ballot (0.005), california (0.004), voter\_fraud (0.004), covid (0.004), voter (0.004)\\
7 & black\_life (0.018), god\_bless (0.008), qanon (0.008), hillary\_clinton (0.008), wearing\_mask (0.007), black (0.007), matter (0.006), life\_matter (0.005), life (0.005), kyle\_rittenhouse (0.005)\\
8 & year\_old (0.008), meme (0.006), cdc (0.005), wear\_mask (0.005), mask (0.005), press (0.005), national\_guard (0.005), lockdown (0.005), general\_flynn (0.004), big\_pharma (0.004)\\
\toprule
\textbf{Topic} & {\bf Words per topic for baseline subverses}\\
\midrule
1 & year\_old (0.024), civil\_war (0.012), california (0.008), old (0.007), last\_night (0.006), year (0.005), question (0.005), san\_francisco (0.005), candidate (0.005), democrat\_party (0.005)\\
2 & people (0.006), joe\_biden (0.006), white (0.005), biden (0.005), trump (0.004), get (0.004), would (0.004), know (0.004), think (0.004), say (0.004)\\
3 & ben\_garrison (0.007), kyle\_rittenhouse (0.005), state (0.005), ballot (0.005), chicago (0.004), deep\_state (0.004), poll (0.004), banned (0.004), check (0.004), good (0.003)\\
4 & supreme\_court (0.016), new\_york (0.015), social\_medium (0.009), cnn (0.007), make\_sure (0.007), court (0.007), tucker\_carlson (0.006), supreme (0.005), piece\_shit (0.005), york (0.005)\\
5 & ghislaine\_maxwell (0.008), fuck (0.006), nigger (0.005), lol (0.005), defund\_police (0.005), china (0.005), faggot (0.005), right (0.004), long\_time (0.004), useful\_idiot (0.004)\\
6 & year\_ago (0.01), trump (0.008), election (0.008), trump\_supporter (0.007), thanks (0.006), coronavirus (0.006), united\_state (0.006), biden (0.006), virus (0.005), anti\_white (0.005)\\
7 & kamala\_harris (0.016), wear\_mask (0.016), deleted (0.01), wear (0.01), make\_sense (0.009), mask (0.009), bill\_gate (0.008), white\_supremacist (0.008), seattle (0.007), law\_enforcement (0.007)\\
8 & president\_trump (0.023), yes (0.013), donald\_trump (0.012), critical\_race (0.01), theory (0.007), conspiracy\_theory (0.007), president (0.007), blm\_antifa (0.007), trump (0.007), donald (0.006)\\
9 & black\_life (0.028), george\_floyd (0.014), black (0.01), matter (0.01), life\_matter (0.009), life (0.009), tranni (0.009), south\_africa (0.008), george (0.008), mail\_ballot (0.007)\\
10 & jew (0.011), right\_wing (0.008), voter (0.008), someone\_else (0.006), interview (0.006), illegal\_alien (0.006), evidence (0.005), public\_school (0.005), meme (0.005), alien (0.005)\\
\toprule
\end{tabular}
\reduce\reduce
\caption{LDA analysis of \greatawakening and baseline subverses.}\label{tbl:topics}
\reduce\reduce
\end{table*}

\subsection{User Registrations}
We also analyze all users' registration dates to understand when they registered a new account on Voat.
Since 2015, online press outlets have reported that communities banned from Reddit often migrate to Voat~\cite{herman2015whatisvoat,wp2018banq,emerson2020whatisvoat}; thus, we investigate whether Voat user registrations increase when Reddit bans communities.
During our data collection period, over 15K users posted a submission or a comment on the subverses.
Also, $13.16\%$ (2K) of these users deactivated their account, or their account was deleted by Voat, due to 404 errors our crawler received from Voat's API.

Figure~\ref{fig:qanon_registrations} and Figure~\ref{fig:base_registrations} plot the number of registered users engaged on \greatawakening and baseline subverses, respectively, per month.
On \greatawakening, the average monthly registration is 4.1, 38.1, 22.75, 28, 125.9, 69.1, and 75 for 2014, 2015, 2016, 2017, 2018, 2019, and 2020, respectively.
Similarly, every month 8.6, 118.1, 50.9, 65.5, 179.6, 142.1, and 200 new user registrations were made, on average, in the baseline subverses.
Over $17.6\%$ (2.3K) unique users registered on Voat in September 2018 only, i.e., the month Reddit banned many QAnon-related subreddits~\cite{wired2018qanon,wp2018banq,stuff2018bansq}.
We also observe another spike in user registration in both communities between June and July 2015, probably due to Reddit banning hate-focused subreddits~\cite{wired2015coon,motherboard2015nigger,verge2015fph}.

Although our dataset might not represent Voat's user base as a whole, it indicates the dates users decided to join the platform.
Looking only at users engaged in baseline subverses (Figure~\ref{fig:base_registrations}), we confirm that Voat received a high volume of new user registrations close to when Reddit banned hateful subreddits and QAnon related subreddits.
Future work, in conjunction with Reddit data, might help shed more light on the effect of Reddit deplatforming and consequent user migration.

\subsection{Take Aways}
Overall, this section answers our RQ1, i.e., {\em how active is the QAnon movement on Voat?}
The most popular QAnon-focused subverse, \greatawakening, attracts many more submissions than the baseline subverses, despite the latter are among the top 10 most popular on the platform for number of subscribers. 
Also, \greatawakening has always more than 50 new submissions per day, with that number steadily and staying above 100 new submissions per day since September 25, 2020.
Whereas the number of daily submissions stays in the same margins for the baseline subverses, except for \askvoat, where we observe a decline in posting activity.

Moreover, both communities' audiences tend to comment on and upvote the submissions and comments they see in the subverse.
Also, the audience of \greatawakening consumes information from just a handful of users, while top submitters and commenters seem to overlap between \greatawakening and the baseline subverses.
Finally, we show that new user registrations peaked after Reddit banned hateful and QAnon subverses in June 2015 and in September 2018, respectively.

\begin{table*}[t]
\centering
\resizebox{1\textwidth}{!}{
\smallskip\begin{tabular}{l r r | l r r | l r r | l r r}
\toprule
\multicolumn{6}{c|}{\textbf{\greatawakening}} & \multicolumn{6}{c}{\textbf{Baseline subverses}}\\
\toprule
\textbf{Named Entity} & \textbf{\#Posts} & ($\%$) & \textbf{Entity Label} & \textbf{\#Posts} & \textbf{($\%$)} & \textbf{Named Entity} & \textbf{\#Posts} & ($\%$) & \textbf{Entity Label} & \textbf{\#Posts} & \textbf{($\%$)} \\
\midrule
Trump (PERSON) & 5,953 & 3.94 & ORG & 69,056 & 45.75 & one (CARDINAL)& 7,621 & 2.13 & ORG & 61,474 & 17.21\\
one (CARDINAL)& 3,623 & 2.40 & PERSON & 61,556 & 40.78 & jews (NORP) & 6,385 & 1.78  & PERSON & 58,383 & 16.34\\
first (ORDINAL)& 2,670 & 1.76 & GPE & 31,286 & 20.74 & first (ORDINAL) & 5,401 & 1.51 & NORP & 40,808 & 11.42\\
US (GPE)& 2,022 & 1.34 & DATE & 29,496 & 19.54 & Jews (NORP)& 4,804 & 1.34 & GPE & 37,947 & 10.62\\
Biden (PERSON) & 2,009 & 1.33 & CARDINAL & 26,155 & 17.32 & Trump (PERSON)& 4,331 & 1.21 & CARDINAL & 35,050 & 9.81\\
America (GPE)& 1,733 & 1.14 & NORP & 20,665 & 13.69 & US (GPE)& 3,571 & 0.99 & DATE & 34,657 & 9.70\\
China (GPE)& 1,660 & 1.09 & WORK\_OF\_ART & 5,481 & 3.63 & two (CARDINAL)& 3,293 & 0.92 & ORDINAL & 9,043 & 2.53\\
two (CARDINAL)& 1,526 & 1.01 & ORDINAL & 5,225 & 3.46 & America (GPE)& 3,142 & 0.88 & LOC & 8,060 & 2.25\\
American (NORP)& 1,505 & 0.99 & TIME & 4,126 & 2.73 & jewish (NORP)& 2,948 & 0.82 & WORK\_OF\_ART & 7,320 & 2.05\\
FBI (ORG) & 1,447 & 0.95 & LOC & 3,900 & 2.58 & Jew (NORP)& 2,305 & 0.64 & PERCENT & 5,816 & 1.62\\
\toprule
\end{tabular}}
\reduce\reduce
\caption{Top ten named entities and entity labels mentioned in \greatawakening and all baseline subverses.}
\reduce\reduce
\label{tbl:named_entities}
\end{table*}

\section{Narrative Analysis}\label{sec:contentanalysis}

In this section, we shed light on the QAnon movement's narrative on Voat, aiming to answer RQ2.
We explore the topics that \greatawakening discusses, and detect the most popular entities they mention using entity detection.
Finally, we use word embeddings and graph representations to visualize keywords most similar to ``qanon.''
We warn readers that some of the content presented and discussed in this section may be disturbing.

\subsection{Topics}\label{sec:topics}
We analyze the most prominent topics on our dataset by running Latent Dirichlet Allocation (LDA)~\cite{blei2003latent} on the text included in both the title and the body of all submissions as well as their comments.
For every post, we remove all the URLs, stop words (e.g., ``like,'' ``to,'' ``and''), and formatting characters, e.g., \textbackslash n, \textbackslash r.
Then, we tokenize each post and analyze it to detect bigrams and include them in our corpus. 
We do this as previous work suggests that bigrams improve the accuracy of topic modeling~\cite{wang2012baselines}.
Last, we create a term-frequency inverse-document frequency (TF-IDF) array to fit an LDA model.
We use a TF-IDF array instead of the default LDA approach as TF-IDF statistically measures every word's importance within the overall collection of words. More importantly, previous work suggests it yields more accurate topics~\cite{mehrotra2013improving}.
We use guidelines from Li~\cite{li2018lda} to build the LDA model.

To measure the appropriate number of topics for our model, we calculate the coherence value (c\_v) of the model for topic numbers between 4 and 20 with step 1.
For \greatawakening, the best coherence score is 0.385 with 8 topics, while for the baseline subverses, the highest coherence score is 0.357 with 10 topics.

In Table~\ref{tbl:topics}, we list the words per topic, along with their weights, discussed on both \greatawakening and the baseline subverses.
For \greatawakening,  users tend to discuss the US Presidential Elections, as suggested by words like ``trump,'' ``election,'' ``biden,'' and ``vote'' across many topics.
Users also refer to ``mail\_ballot'' and ``voter\_fraud,'' (topic 6) along with ``posts'' from ``potus'' about ``cnn'' ``fake\_news'' (topic 4) 
There are also discussions about the COVID-19 pandemic: ``wear\_mask,'' ``lockdown,'' and ``big\_pharma'' (topic 8).
We also find a topic about the ``Black Lives Matter'' movement, ``black\_life,'' ``black,'' ``life\_matter,'' (topic 7).
As for baseline subverses, we once again find topics including elections, coronavirus, and Black Lives Matter, but with even more frequent hateful words such as ``fuck,'' ``nigger,'' ``tranni,'' ``faggot,'' etc.

Overall, our topic detection analysis shows that discussions on \greatawakening revolve around Trump and political matters, where baseline subverses feature news, along with hateful and controversial words.
We will further analyze toxicity in Section~\ref{sec:toxicity}.

\subsection{Named Entities}
While topic modeling gives us an idea of \emph{what} is being discussed, to get an understanding of \emph{who} is being discussed, we extract the named entities used in our communities of interest.
We do so to understand who conspiracies focus on and better define the narratives they might be pushing.

To obtain the named entities mentioned in each post, we use the en\_core\_web\_lg (v2.3) model from the SpaCy library~\cite{spacy}. 
We select this specific model over alternatives, e.g., MonkeyLearn, %
since, to the best of our knowledge, it is trained on the largest training set. 
Moreover, previous work~\cite{jiang2016evaluating} ranks it as the second most accurate method for recognizing named entities in text, with the first being Stanford NER.
We choose en\_core\_web\_lg over Stanford NER as it detects dates more accurately.  
The model uses millions of online news outlet articles, blogs, and comments from various social networks to detect and extract various entities from text.
Crucially for our purposes, it also provides an entity category label in addition to the entity itself.
For example, the entity category for Donald Trump is ``person.'' 
The different categories range from celebrities to nationalities, products, and events.\footnote{See \url{https://spacy.io/api/annotation\#named-entities} for the full list of labels.} 

In Table~\ref{tbl:named_entities}, we list the ten most popular named entities and categories from \greatawakening and all the baseline subverses.
Note that a post may mention an entity more than once. 
Therefore, we only report the number of posts that mention an entity at least once. 
Unsurprisingly, considering his central role in the QAnon conspiracy, ``Donald Trump'' is the most popular named entity on \greatawakening with almost 6K posts ($3.94\%$) mentioning him.
Other popular entities mentioned in \greatawakening include ``US'' ($1.34\%$), ``Biden'' ($1.33\%$), ``America'' ($1.14\%$), ``China'' ($1.09\%$), and ``FBI''' ($0.95\%$).
The most popular category is organizations ($45.75\%$), followed by people ($40.78\%$). %
Other popular labels include nationalities, religious, or political groups (NORP, $13.69\%$), books, songs, and movies (WORK\_OF\_ART, $3.63\%$), and times ($2.73\%$). 
In comparison, the most popular named entities mentioned in the baseline subverses are ``jews'' ($2.58\%$), ``Trump'' ($1.21\%$), ``America'' ($0.88\%$), and ``jewish'' ($0.82\%$).
The most popular labels are organizations ($17.21\%$), people ($16.34\%$), and nationalities, religious, or political groups ($11.42\%$). 

Overall, this suggests that discussions within these communities are related to US happenings and events, politics, and established organizations and institutions.
Baseline subverses focus mostly on nationalities, and religious or political groups, while \greatawakening discussions focus on the US, Donald Trump, and the US Presidential elections.

\begin{figure*}[t!]
\centering
    \includegraphics[width=1\textwidth]{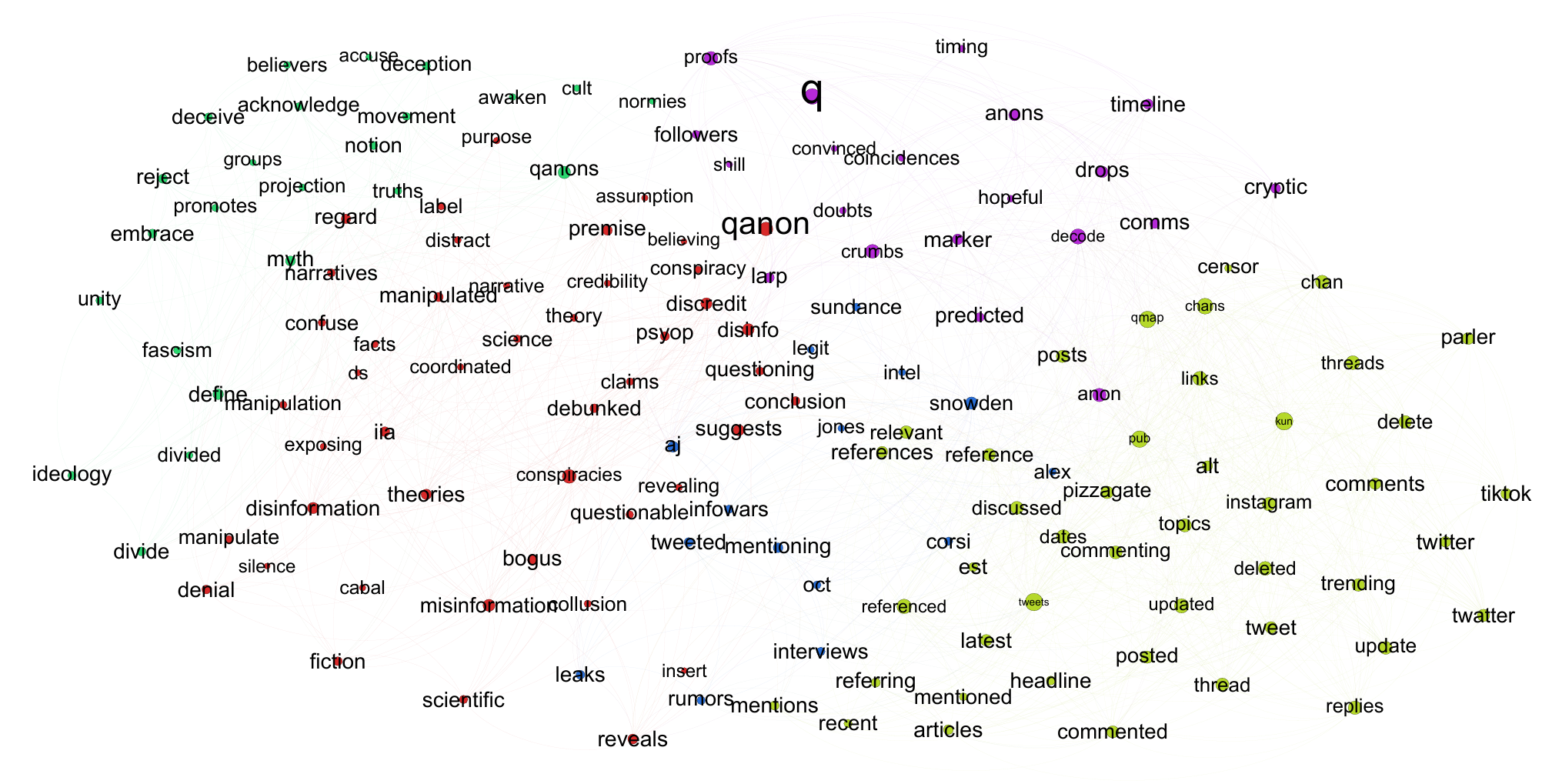}
    \caption{Graph representation of the words associated with the term ``qanon'' on Voat. 
}
    \label{fig:qanon_graph}
\end{figure*}

\subsection{Text Analysis}\label{sec:w2v}
\descr{Word Embeddings.} To assess how different words are interconnected with popular QAnon specific keywords (e.g., ``qanon''), we analyze our \greatawakening dataset using word2vec, a two-layer neural network that generates word representations as embedded vectors~\cite{mikolov2013efficient}. 
A word2vec model takes a large input corpus of text and maps each word in the corpus to a generated multidimensional vector space, yielding a \emph{word embedding}. 
Words that are used in similar contexts tend to have similar vectors in the generated vector space.

To clean the QAnon posts before training the model, we follow a similar methodology as for the topic modeling presented in Section~\ref{sec:topics}.
We train the word2vec model using a context window (which defines the maximum distance between the current word and predicted words when generating the embedding) of 7, as suggested by~\cite{li2017data}.
We limit the corpus to words that appear at least 50 times due to our dataset's small size. %
Finally, we train the word2vec model with 8 iterations (epochs) as, on small corpora like ours, epochs between 5 and 15 epochs are suggested to provide the best results~\cite{mikolov2013efficient,mikolov2013distributed}.
(Choosing more epochs than 8 makes our model overfit and minimizes the word vocabulary, e.g., removing QAnon-specific keywords like ``qanon.'')
After training, our model includes a 5.6K word vocabulary.

\descr{QAnon similar keywords.}
Next, we find the top ten most similar words to ``qanon'' and ``q'' according to the model; see Table~\ref{tbl:qanon_similar}.
We see that ``qanon'' is linked to words like ``conspiracy,'' ``theories,'' ``movement,'' and ``pizzagate.''
The term ``q'' seems to be closely related to Q's activity and the research the community does to decode his cryptic messages as evident to ``drops,'' which refers to the posts that Q leaves as breadcrumbs of information for adherents of the conspiracy to decode.
These drops often hint at ``psyops'', the alleged psychological operations the deep state and governments deploy to control society.
Interestingly, the term ``larp,'' an acronym for ``Live Action Role Playing,'' is sometimes used in a derogatory fashion to imply that Q is just a troll playing a game.
This indicates that even on a community devoted to the QAnon conspiracy, there is at least some degree of pushback or dissent within the user base.
We use graph representations to analyze this finding below.

\begin{table}[t]
\centering
\small
\smallskip\begin{tabular}{l r | l r }
\toprule
\multicolumn{2}{c}{\textbf{``qanon''}}  & \multicolumn{2}{c}{\textbf{``q''}}   \\
\midrule
\textbf{Word} & \textbf{Cos.~Similarity} & \textbf{Word} & \textbf{Cos.~Similarity} \\
\midrule
conspiracy & 0.636 & anons & 0.679 \\
theories & 0.582 & larp & 0.594 \\
q & 0.579 & qanon & 0.579 \\
movement & 0.570 & drops & 0.570 \\
followers & 0.561 & proofs & 0.557 \\
conspiracies & 0.549 & cryptic & 0.545 \\
tweets & 0.547 & psyop & 0.531 \\
aj & 0.545 & posts & 0.529 \\
qanons & 0.544 & anon & 0.526\\
discredit & 0.538 & crumbs & 0.524 \\
\toprule
\end{tabular}
\reduce\reduce
\caption{Top ten similar words to the term ``qanon'' and ``q'' and their respective cosine similarity.}
\reduce\reduce\reduce
\label{tbl:qanon_similar}
\end{table}

\begin{figure}[t]
    \centering
    \includegraphics[width=0.9\columnwidth]{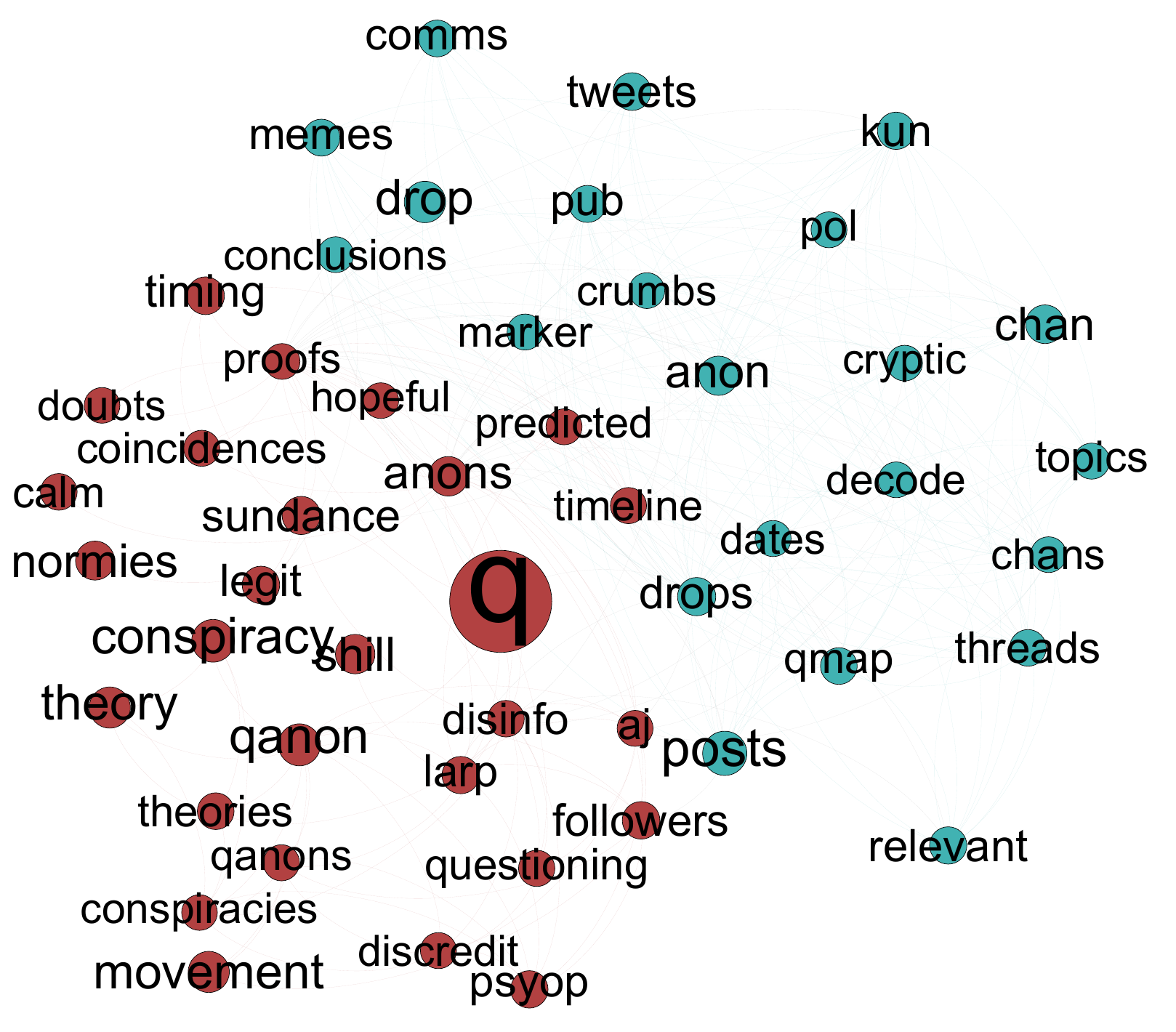}
    \reduce
    \caption{Graph representation of the words associated with the term ``q'' on Voat.}
    \reduce
    \label{fig:q_graph}
\end{figure}

\descr{Graph representations.}
We follow the methodology from~\cite{zannettou2020quantitative} to visualize topics within the word embeddings.
Specifically, we transform the embeddings into a graph, where nodes are words and edges are weighted by the cosine similarity between the learned vectors of the nodes the edge connects.
We perform community detection~\cite{blondel2008fast} on the resulting graph to gain new insights into the high-level topics that groups of words form.

\descr{Visualization.} 
Figure~\ref{fig:qanon_graph} shows the two-hop ego network centered around the word ``qanon.'' %
Figure~\ref{fig:q_graph} depicts a graph centered around the word Q.
To improve readability (since our graph transformation results in a fully connected network), we remove all edges with a cosine similarity less than 0.6.
We further color each node based on the community it belongs to.
Finally, we apply the ForceAtlas2 algorithm~\cite{jacomy2014forceatlas2}, which considers the edges' weight when laying out the nodes in the 2-dimensional space, before producing the final visualization.

\descr{Remarks.} Taking into account how communities form distinct themes, and that nodes' proximity implies contextual similarity, Figure~\ref{fig:qanon_graph} shows that the ``qanon'' community (red) is very close to the green (far left on the figure) one, which seems to be discussing the movement itself (``qanons,'' ``cult,'' ``fascism,'' ``believers,'' ``movement,'').
The small blue community in the middle of the figure discusses ``leaks'' and ``interviews'' from Edward ``snowden.''
Next, the purple community is focused on Q's activity and the posts he drops (``q,'' ``drops,'' ``timeline,'' ``decode,'' ``cryptic'').
In the yellow community (far right on the figure), we come across the QAnon predecessor ``pizzagate,'' Q drop aggregators (e.g., ``qmap,'' which was recently shut down~\cite{qmapdown}), and other social media platforms (8``kun'', 4``chan,'' ``twitter,'' ``instagram,'' and ``parler'').

Focusing on the conspiracy theory's originator, Figure~\ref{fig:q_graph} plots the discussion around Q.
Interestingly, the community of ``q'' (red) has words like ``larp,'' ``disinfo,'' ``doubts,'' and ``shill'' (a term used for someone that might be hired by the government pretending to agree with a conspiracy) in close proximity of Q.
On the other hand, we find terms like ``followers'' and ``aj'' (a term used to describe a man as supportive and perfect).
This plot strengthens the hypothesis that although the community is devoted to the QAnon movement, at the same time, there might be signs of chasm with regards to what the users on \greatawakening think of Q.
Finally, the blue community discusses Q's ``cryptic'' ``drops,'' and various social networks like 4``chan,'' 8``kun,'' and ``qmap'' aggregation site that archives Q's posts.

\subsection{Take Aways}
The analysis presented in this section allows us to identify and visualize the narratives around QAnon discussion (RQ2).
We show that the QAnon community discusses online social media, political matters, and world events.
Additionally, the main topic of conversation is Donald Trump and the US overall, and entities discussed are most typically organizations and individuals.
These findings confirm that, regardless of the conspiracy theory's particular components, Trump's role in the conspiracy, e.g., as the alleged leader in the war against the deep state, is central.

Finally, our structural analysis of word embedding similarities provides some high-level discussion topics within the community.
For example, we find that the term ``larp,'' an oft used criticism of Q implying he is merely playing a game, is often used in the same context as discussion of ``q'' himself.
This is an indicator that adherents are well aware of criticisms of their information source, and perhaps some dissent within the community itself.
Additionally, we see that the movement is well embedded across the Web, with external q-drop aggregators (e.g., qmap) and social media platforms are commonly discussed along with Q.

\begin{figure*}[t!]
\center
\subfigure[]{\includegraphics[width=0.33\textwidth]{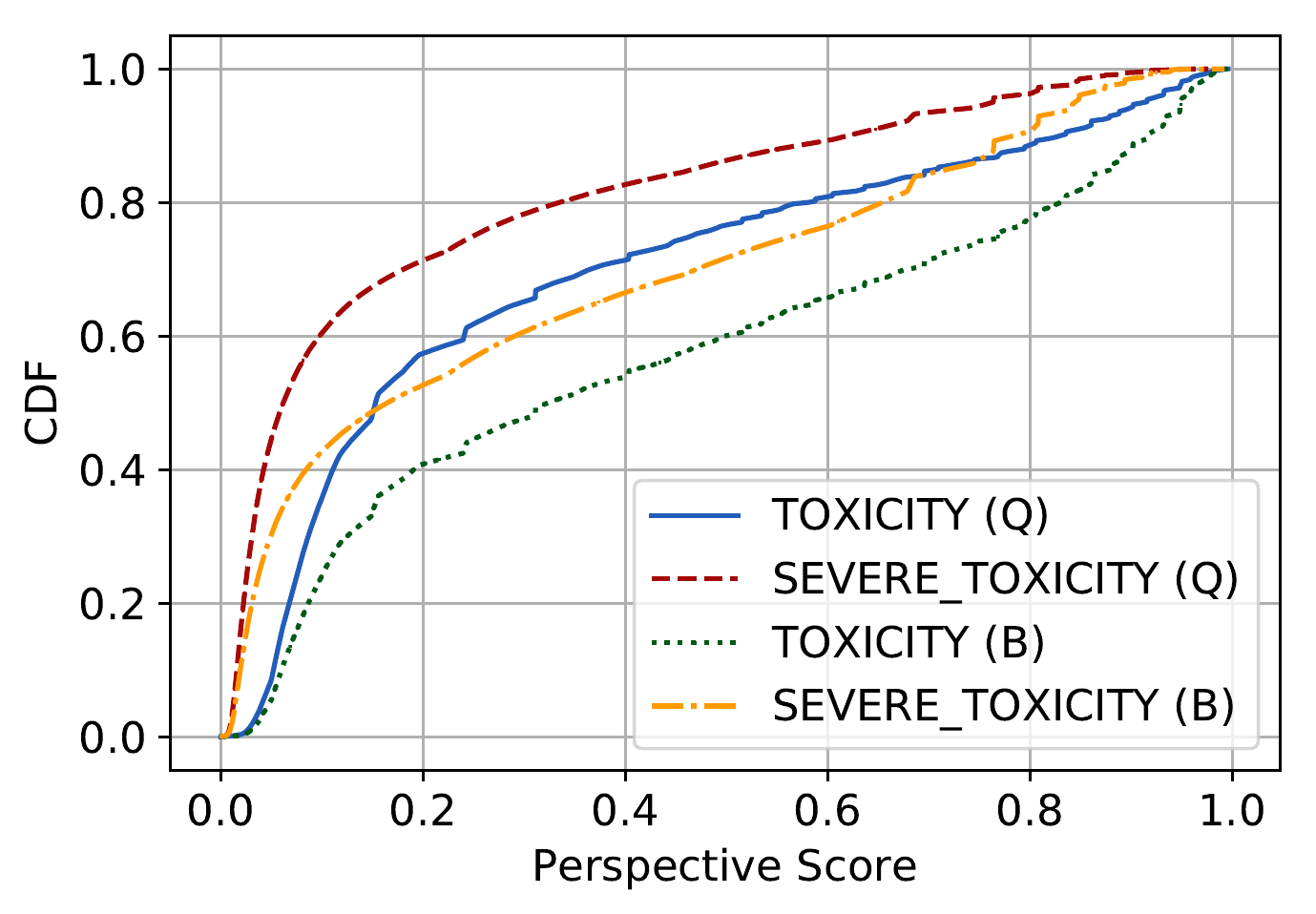}\label{fig:cdf_toxicity}}
\subfigure[]{\includegraphics[width=0.33\textwidth]{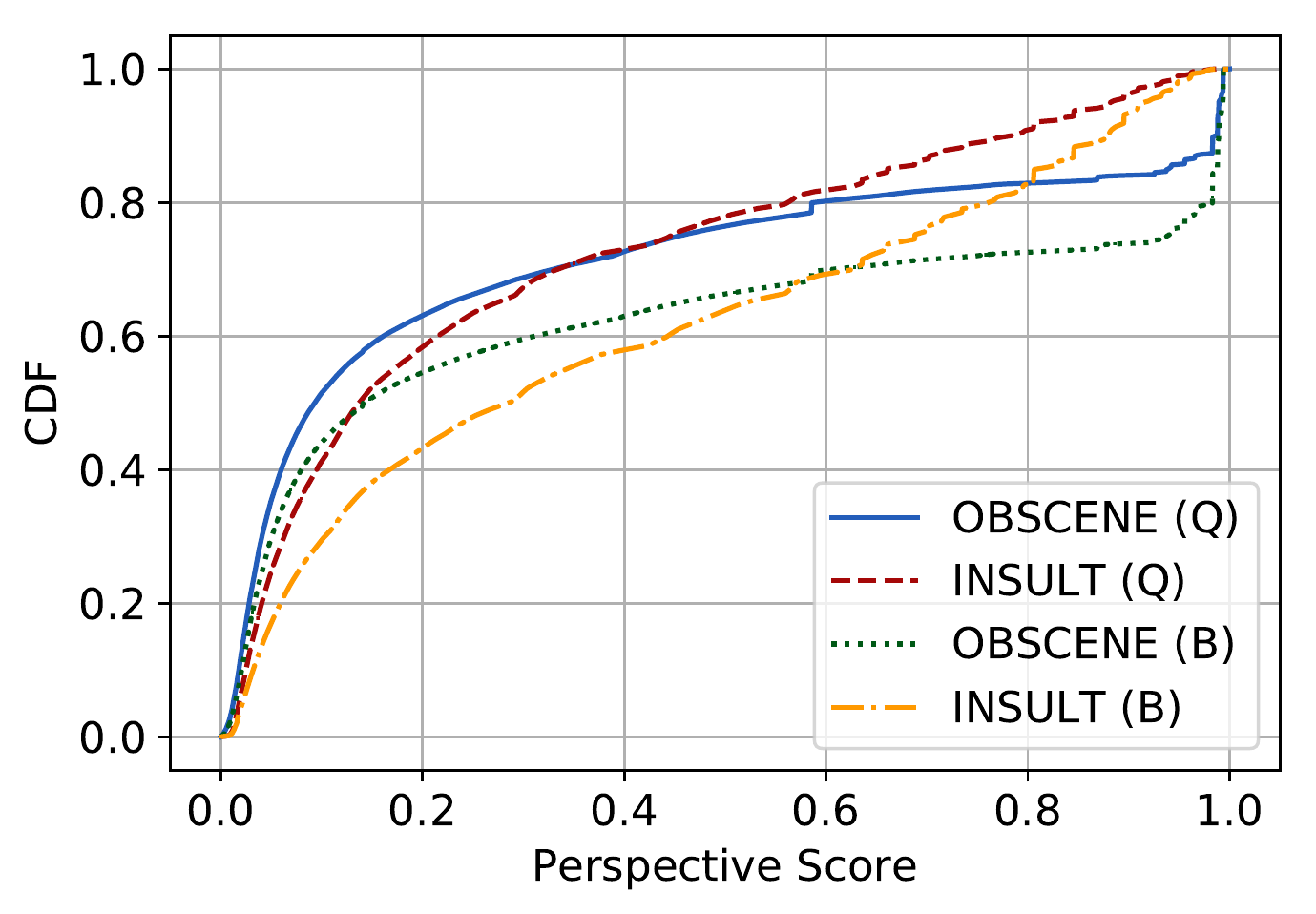}\label{fig:cdf_insult}}
\subfigure[]{\includegraphics[width=0.33\textwidth]{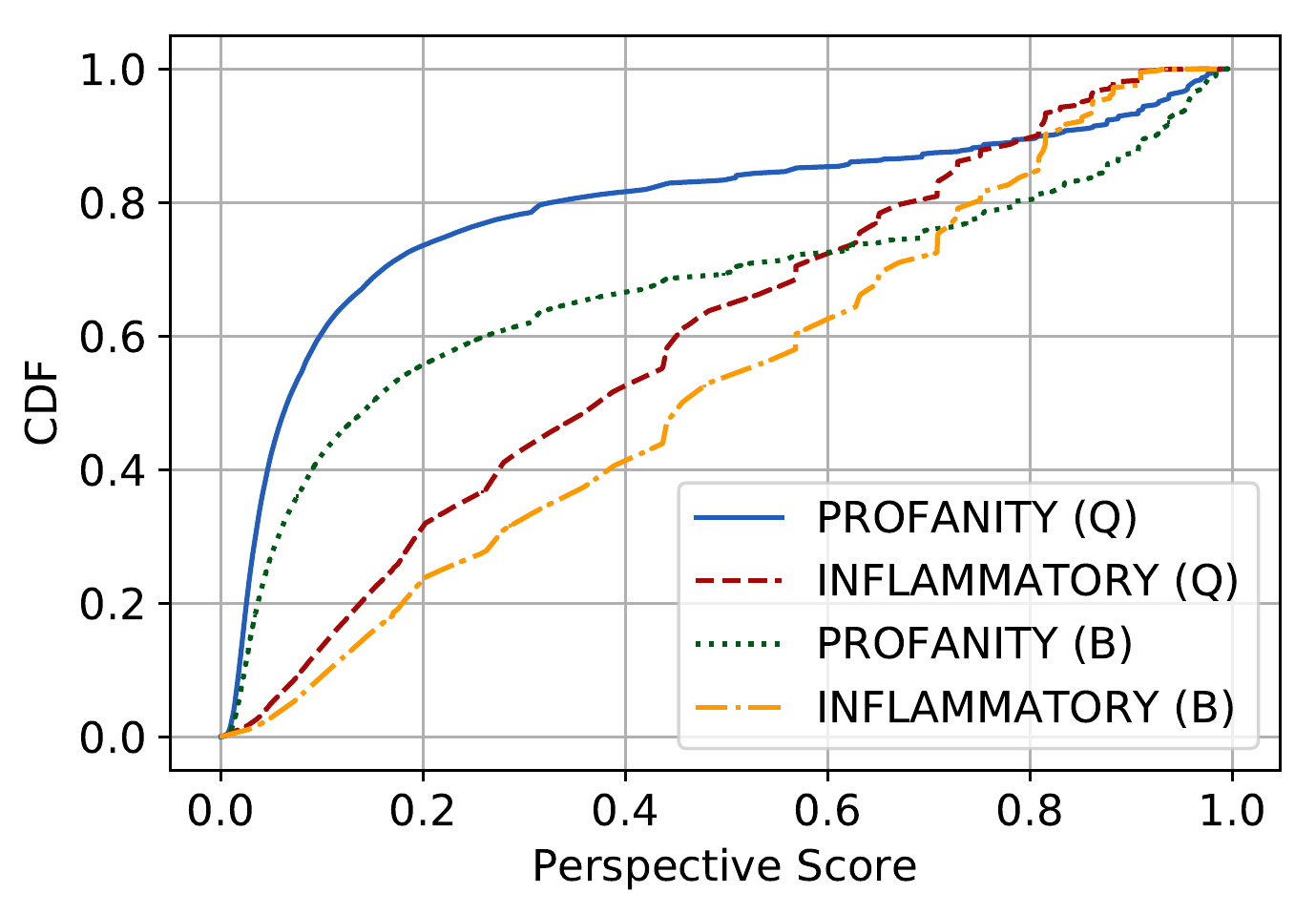}\label{fig:cdf_inflammatory}}
\reduce\reduce
  \caption{CDF of the Perspective Scores related to how toxic, severely toxic, obscene, insulting, profane, and inflammatory a post is for the \greatawakening (Q) and baseline (B) subverses.}
\label{fig:cdf_perspective_scores}
\end{figure*}

\section{Toxicity Analysis}\label{sec:toxicity}
In this section, we analyze the toxicity of the \greatawakening community, compared to the general discussion subverses.

Motivated by our earlier findings suggesting that toxicity, hate, and racism exist in all subverses of our dataset, we measure how toxic, obscene, insulting, profane, and inflammatory each post is.
To do so, we use Google's Perspective API~\cite{jigsaw2018perspective}.
We choose this tool, similar to prior work~\cite{papasavva2020raiders}, as other methods mostly use short texts (tweets) for training~\cite{davidson2017automated}, whereas Google's Perspective API is trained on crowdsourced annotations and comments with no restriction in character length, similar to Voat posts.
We also acknowledge the limitations of the API; namely, false-negative results due to misspelled words~\cite{hosseini2017deceiving} and bias against African American English written posts~\cite{sap2019risk}.
However, we do not take the scores at face value but use them to compare the differences between QAnon-related and baseline posts on Voat. %

We rely on six models to annotate posts from all subverses: {\em toxicity}, {\em severe\_toxicity}, {\em obscene}, {\em insult}, {\em profanity}, and {\em inflammatory}.\footnote{See~\url{https://github.com/conversationai/perspectiveapi/blob/master/2-api/model-cards/English/toxicity.md} for the details of each model.}
Note that all methods provide scores (0 to 1) for textual posts. %
Therefore, we do not have scores for $4.8\%$ (24.6K) of the posts in our dataset since they only contain links or images but no text.
In Figure~\ref{fig:cdf_perspective_scores}, we plot the CDF of the scores for each  model.
The baseline subverses (B in Figure~\ref{fig:cdf_toxicity}) exhibit higher levels of {\em toxicity} and {\em severe\_toxicity}, compared to \greatawakening (Q in the figure).
Specifically, $39.9\%$ and $28.2\%$ of the baseline posts have, respectively, {\em toxicity} and {\em severe\_toxicity} scores greater than 0.5, while only $23.3\%$ and $13.7\%$ of the QAnon posts have these scores greater than 0.5.
We observe similar trends for the other models, with the baseline subverses always scoring higher than \greatawakening.
Overall, $33.6\%$ and $36\%$ of the baseline subverses' posts have an {\em obscene} and {\em insult} score greater than 0.5, respectively (Figure~\ref{fig:cdf_insult}), and $33.6\%$ for {\em profanity} and $46\%$ for {\em inflammatory} (Figure~\ref{fig:cdf_inflammatory}).
For all six models, the percentage of the QAnon posts that have perspective score greater than 0.5 is at least $10\%$ smaller than the general discussion posts.
Last, we use two-sample KS test to check for statistically significant differences between all the distributions in Figure~\ref{fig:cdf_perspective_scores} and find them ($p < 0.01$). 

\descr{Remarks.} %
Although the QAnon community's content exhibits some levels of toxicity, the movement is not as toxic as other discussions on the platform. 
We believe this not to be entirely surprising as the community seems to be more focused on the conspiracy aspects of world events, politics, and Donald Trump, %
while racist or hateful agendas might more vigorously characterize Voat as a whole, or at least the popular general-discussion subverses in our baseline.
In other words, toxicity in the discussions seems to target the so-called ``deep-state,'' the puppet masters, and the pedophile ring members.
Whereas baseline subverses like \news and \politics are likely to include inflammatory discussions between users with contradicting opinions, or comment on world events from a racist/hateful standpoints.

Interestingly, the baseline subverses' level of toxicity appears to be similar to that of 4chan's \dspol, which is measured in~\cite{papasavva2020raiders}.
In particular, we find that the percentage of posts that get scores above 0.5, across all models are very similar on \dspol and our four baseline subverses.
Considering that \dspol is broadly considered to be a highly toxic place~\cite{hine2016kek}, this suggests that Voat is too.

\section{Related Work}\label{sec:relatedwork}
In this section, we review previous work on QAnon and Voat.

\descr{Qualitative work on QAnon.}
Prooijen~\cite{Prooijen2018ThePO} studies why people believe in conspiracy theories like QAnon, arguing that their beliefs are not necessarily pathological or novel and can be followed by individuals who behave relatively normally.
The author explains that, typically, individuals follow more than one conspiracy theory, as also discussed by Goertzel~\cite{goertzel1994belief}, %
and they believe that nothing happens coincidentally. 
At their core, conspiracy theories reinforce the idea that hostile or secret machinations permeate all social layers, thus forging an appealing account of events for the individuals that seek ``explanations,'' especially after experiencing anxiety and uncertainty due to societal events that traumatized them.

Sternisko et al.~\cite{sternisko2020dark} %
argue that conspiracy theories pose a real threat to democracies, as governments and media might start or amplify them to benefit their political agendas and interests.
Schabes~\cite{schabes2020birtherism} stresses that social networks help conspiracy theories spread faster, which threatens individual autonomy and public safety, enforces political polarization, and harms trust in government and media.
Rutschman~\cite{santos2020mapping} explains that misinformation spread by the QAnon movement can be dangerous to individuals, e.g., 
claiming that drinking chlorine dioxide prevents COVID-19 infections.
Thomas and Zhang~\cite{thomasid2020} explain that small groups of engaged conspiracists, like QAnon followers, can potentially influence recommendation algorithms to expose new, unsuspecting users to their beliefs.
The same study notes that conspiracy theories often include information from legitimate sources or official documents framed with misleading and conspiratorial explanations to events, which creates illusions and further complicates moderation efforts against conspiratorial content.

\descr{Quantitative work on QAnon.}
McQuillan et al.~\cite{mcquillan2020cultural} collect 81M tweets related to COVID-19 between January and May 2020, finding that the QAnon movement not only has grown throughout the pandemic but also that its content has reached more mainstream groups.
In fact, the Twitter QAnon community almost doubled in size within two months.
Darwish~\cite{darwish2018kavanaugh} gathers 23M tweets related to US Supreme Court judge Brett Kavanaugh for 3 days and 4 days in September and October 2018, respectively.
They find that the hashtags \#QAnon and \#WWG1WGA (Where We Go One We Go All) are in the top 6 hashtags in their dataset.
Chowdhury et al.~\cite{chowdhury2020twitter} identify 2.4M accounts suspended from Twitter and collect 1M tweets, performing a retrospective analysis to characterize the accounts and their behavioral activities.
They observe that politically motivated users consistently and successfully spread controversial and political conspiracies over time, including the QAnon conspiracy.

Faddoul et al.~\cite{faddoul2020longitudinal} collect the top-recommended YouTube videos from 1,080 YouTube channels between October 2018 and February 2020.
In total, they analyze more than 8M recommendations from YouTube's watch-next algorithm and use 500 videos labeled as ``conspiratory'' to train a classifier to detect conspiracy-related videos with $78\%$ precision.
Using TF-IDF, they also find that, within the top 15 discriminating words in the snippet of the training set videos, the term ``qanon'' ranks third.
Also, QAnon-related videos belong to one of the three top topics identified by an unsupervised topic modeling algorithm. 
The authors conclude that YouTube's recommendation engine might operate as a ``filter bubble.'' %
Recently, Aliapoulios et al.~\cite{aliapoulios2021gospel} collect Q drops archived by six ``aggregation sites'' to study QAnon from Q's perspective, and how links to these sites are shared on platforms like Twitter and Reddit.

\descr{Voat.}
Chandrasekharan et al.~\cite{chandrasekharan2017bag} detect abusive content using data from 4chan, Reddit, MetaFilter, and Voat, and relying on a novel approach called Bag of Communities (BoC).
Part of the Voat data collected for their work originates from \coon, \nigger, and \fph: three communities focused on hate towards groups of individuals with specific body or race characteristics. 
These subverses were created in Voat after Reddit banned the original \rcoon, \rfph, and \rnigger subreddits in 2015~\cite{wired2015coon,motherboard2015nigger,verge2015fph}.
Similarly, Salim et al.~\cite{saleem2017web} use Reddit and Voat's \coon, \fph, and \redpill comments to train a classifier to detect hateful speech.
Khalid and Srinivasan~\cite{khalid2020style} collect 872K comments from \politics, \television, and \travel in an attempt to detect distinguishable linguistic style across various communities; more specifically, they compare the features of Voat comments to Reddit and 4chan comments and train a classifier to predict the origin of a comment based on its style and content.
Finally, Popova~\cite{popova2019reading} uses data from \deep and mrdeepfakes.com, finding that pornographic deepfakes are often created for circulation and enjoyment within the community.
Note that both the mrdeepfakes.com and the subverse \deep were created after Reddit banned the subreddit \rdeep in 2018~\cite{verge2018deep,daily2018deep}.

\descr{Remarks.}  Previous quantitative work related to QAnon has mostly focused on Twitter~\cite{chowdhury2020twitter,mcquillan2020cultural}, while ours does so on Voat.
Overall, our paper presents, to the best of our knowledge, the first characterization of the QAnon community on Voat.
Some of our findings are aligned with those from previous studies, e.g., a steady increase in posting activity on \greatawakening, somewhat similar to~\cite{mcquillan2020cultural}, which finds that the QAnon movement on Twitter increased in size over their collection period.

\section{Conclusion}\label{sec:conclusion}
This work presented a first characterization of the QAnon movement on the social media aggregator site Voat. 
We collected over 510K posts from five subverses: \greatawakening, the largest QAnon-related subverse, as well as a baseline consisting of the four most active subverses, \news, \politics, \funny, and \askvoat.

We showed that users on both the QAnon and baseline subverses tend to be engaged. 
However, the audience of \greatawakening consumes data from just a handful of content creators responsible for over $72.8\%$ of the total submissions in the community.
The \greatawakening subverse had a peak in registration activity shortly after Reddit banned QAnon related communities in September 2018.
Using topic modeling techniques, we showed that conversations focus on world events, US politics, and Donald Trump. 
We also trained a word2vec model to illustrate the connection of different terms to closely related words, finding that the terms ``qanon'' and ``q'' are closely related to other conspiracy theories like Pizzagate, other social networking platforms, the so-called deep-state, and ``research'' activities the community performs to decode Q's cryptic posts.
Finally, toxicity scores from Google's Perspective API show that posts in \greatawakening are \emph{less} toxic than those on popular general-discussion (baseline) subverses.

Although this paper represents the first large-scale study of the QAnon movement on Voat, it is far from comprehensive, and numerous questions about the movement remain, leaving several directions for future work.
First, while this paper focused on Voat, the QAnon movement is decidedly multi-platform, and thus we encourage work that examines it from a cross-platform perspective~\cite{aliapoulios2021gospel}.
Next, even though it has only recently entered mainstream discourse, QAnon has a long and still somewhat muddied evolution.
This calls for longitudinal studies that cover a much longer period than that in the present work to get a firm grasp on how the movement has evolved, both in terms of components of the conspiracy as well as user engagement and discussion (e.g., how do adherents react when the predictions in a q-drop do not come to pass).
Finally, we believe that while understanding the movement itself is important, there are real indications that it exhibits cult-like characteristics -- e.g., recovery stories from former adherents~\cite{businessHeWentQAnon} and communities devoted to emotional support for people whose loved ones have become followers\footnote{\url{https://www.reddit.com/r/QAnonCasualties/}} -- it is crucial to understand more about the QAnon \emph{counter-movement}, which might provide insights into the real-world impact of the spread of dangerous conspiracy theories as well as devising mitigation strategies.

\descr{Acknowledgments.} This project was partly funded by the UK EPSRC grant EP/S022503/1, which supports the Center for Doctoral Training in Cybersecurity delivered by UCL's Departments of Computer Science, Security and Crime Science, and Science, Technology, Engineering and Public Policy.

\small
\bibliographystyle{abbrv}
\bibliography{references}

\begin{thebibliography}{10}

\bibitem{aliapoulios2021gospel}
M.~Aliapoulios, A.~Papasavva, C.~Ballard, E.~De~Cristofaro, G.~Stringhini,
  S.~Zannettou, and J.~Blackburn.
\newblock {The Gospel According to Q: Understanding the QAnon Conspiracy from
  the Perspective of Canonical Information}.
\newblock arXiv:2101.08750, 2021.

\bibitem{bbc2020bansfb}
{BBC}.
\newblock {Facebook bans QAnon conspiracy theory accounts across all
  platforms}.
\newblock \url{https://www.bbc.com/news/world-us-canada-54443878}, 2020.

\bibitem{bbc2020girlfriendarrested}
{BBC}.
\newblock {Jeffrey Epstein ex-girlfriend Ghislaine Maxwell charged in US}.
\newblock \url{https://www.bbc.co.uk/news/world-us-canada-53268218}, 2020.

\bibitem{bbc2020twitterban}
{BBC}.
\newblock {QAnon: Twitter bans accounts linked to conspiracy theory}.
\newblock \url{https://www.bbc.co.uk/news/world-us-canada-53495316}, 2020.

\bibitem{originsqanon}
BBC.
\newblock {QAnon: What is it and where did it come from?}
\newblock \url{https://www.bbc.com/news/53498434}, 2020.

\bibitem{ukqanon}
{BBC}.
\newblock {What's behind the rise of QAnon in the UK?}
\newblock \url{https://www.bbc.com/news/blogs-trending-54065470}, 2020.

\bibitem{blei2003latent}
D.~M. Blei, A.~Y. Ng, and M.~I. Jordan.
\newblock Latent dirichlet allocation.
\newblock {\em JMLR}, 3:993--1022, 2003.

\bibitem{blondel2008fast}
V.~D. Blondel, J.-L. Guillaume, R.~Lambiotte, and E.~Lefebvre.
\newblock Fast unfolding of communities in large networks.
\newblock {\em Journal of Statistical Mechanics: Theory and Experiment},
  2008(10), 2008.

\bibitem{qmapdown}
{Bloomberg}.
\newblock {QAnon website shuts down after NJ man identified as operator}.
\newblock \url{https://bit.ly/2TahE6g}, 2020.

\bibitem{chandrasekharan2017bag}
E.~Chandrasekharan, M.~Samory, A.~Srinivasan, and E.~Gilbert.
\newblock {The Bag of Communities: Identifying Abusive Behavior Online with
  Preexisting Internet Data}.
\newblock In {\em ACM SIGCHI}, 2017.

\bibitem{chowdhury2020twitter}
F.~A. Chowdhury, L.~Allen, M.~Yousuf, and A.~Mueen.
\newblock {On Twitter Purge: A Retrospective Analysis of Suspended Users}.
\newblock In {\em Companion Proceedings of the Web Conference 2020}, 2020.

\bibitem{businessHeWentQAnon}
{CNN}.
\newblock {He Went down the QAnon Rabbit Hole for Almost Two Years. Here's How
  He Got Out}.
\newblock \url{https://cnn.it/3kgfCNJ}, 2020.

\bibitem{daily2018deep}
{DailyDot}.
\newblock {Here's where `deepfakes,' the fake celebrity porn, went after the
  Reddit ban}.
\newblock \url{https://www.dailydot.com/unclick/deepfake-sites-reddit-ban/},
  2018.

\bibitem{darwish2018kavanaugh}
K.~Darwish.
\newblock {To Kavanaugh or not to Kavanaugh: That is the Polarizing Question}.
\newblock arXiv:1810.06687, 2018.

\bibitem{davidson2017automated}
T.~Davidson, D.~Warmsley, M.~Macy, and I.~Weber.
\newblock Automated hate speech detection and the problem of offensive
  language.
\newblock In {\em ICWSM}, 2017.

\bibitem{pizzagate}
{Esquire}.
\newblock {Years After Being Debunked, Interest in Pizzagate Is
  Rising—Again}.
\newblock \url{https://bit.ly/3jilbK9}, 2020.

\bibitem{faddoul2020longitudinal}
M.~Faddoul, G.~Chaslot, and H.~Farid.
\newblock {A Longitudinal Analysis of YouTube's Promotion of Conspiracy
  Videos}.
\newblock arXiv:2003.03318, 2020.

\bibitem{QAnoncongress}
{Forbers}.
\newblock {Congress Will Get Its Second QAnon Supporter, As Boebert Wins
  Colorado House Seat}.
\newblock \url{https://bit.ly/2LyiB7G}, 2020.

\bibitem{goertzel1994belief}
T.~Goertzel.
\newblock Belief in conspiracy theories.
\newblock {\em Political psychology}, 1994.

\bibitem{hine2016kek}
G.~E. Hine, J.~Onaolapo, E.~De~Cristofaro, N.~Kourtellis, I.~Leontiadis,
  R.~Samaras, G.~Stringhini, and J.~Blackburn.
\newblock {Kek, Cucks, and God Emperor Trump: A measurement study of 4chan's
  politically incorrect forum and its effects on the Web}.
\newblock In {\em ICWSM}, 2017.

\bibitem{hosseini2017deceiving}
H.~Hosseini, S.~Kannan, B.~Zhang, and R.~Poovendran.
\newblock {Deceiving Google's Perspective API built for detecting toxic
  comments}.
\newblock arXiv:1702.08138, 2017.

\bibitem{herman2015whatisvoat}
{International Business Times}.
\newblock {Reddit FatPeopleHate Ban Has Users Migrate To Unmoderated Space,
  Complain On Twitter}.
\newblock \url{https://bit.ly/37qQU9R}, 2015.

\bibitem{jacomy2014forceatlas2}
M.~Jacomy, T.~Venturini, S.~Heymann, and M.~Bastian.
\newblock Forceatlas2, a continuous graph layout algorithm for handy network
  visualization designed for the gephi software.
\newblock {\em PloS one}, 2014.

\bibitem{jewsbillgates}
{Jewish Standard}.
\newblock {QAnon conspiracies mirror historic anti-Semitism}.
\newblock
  \url{https://jewishstandard.timesofisrael.com/qanon-conspiracies-mirror-historic-anti-semitism/},
  2020.

\bibitem{jiang2016evaluating}
R.~Jiang, R.~E. Banchs, and H.~Li.
\newblock Evaluating and combining name entity recognition systems.
\newblock In {\em Named Entity Workshop}, 2016.

\bibitem{khalid2020style}
O.~Khalid and P.~Srinivasan.
\newblock {Style Matters! Investigating Linguistic Style in Online
  Communities}.
\newblock In {\em ICWSM}, 2020.

\bibitem{li2017data}
Q.~Li, S.~Shah, X.~Liu, and A.~Nourbakhsh.
\newblock Data sets: Word embeddings learned from tweets and general data.
\newblock In {\em ICWSM}, 2017.

\bibitem{mcquillan2020cultural}
L.~McQuillan, E.~McAweeney, A.~Bargar, and A.~Ruch.
\newblock Cultural convergence: Insights into the behavior of misinformation
  networks on twitter.
\newblock arXiv:2007.03443, 2020.

\bibitem{mehrotra2013improving}
R.~Mehrotra, S.~Sanner, W.~Buntine, and L.~Xie.
\newblock Improving lda topic models for microblogs via tweet pooling and
  automatic labeling.
\newblock In {\em SIGIR}, 2013.

\bibitem{mikolov2013efficient}
T.~Mikolov, K.~Chen, G.~Corrado, and J.~Dean.
\newblock Efficient estimation of word representations in vector space.
\newblock arXiv:1301.3781, 2013.

\bibitem{mikolov2013distributed}
T.~Mikolov, I.~Sutskever, K.~Chen, G.~S. Corrado, and J.~Dean.
\newblock Distributed representations of words and phrases and their
  compositionality.
\newblock In {\em NeurIPS}, 2013.

\bibitem{motherboard2015nigger}
{Motherboard}.
\newblock {Why Reddit Banned Some Racist Subreddits But Kept Others}.
\newblock \url{https://bit.ly/3dHMHjp}, 2015.

\bibitem{nbc2018containQ}
{NBC News}.
\newblock {The far right is struggling to contain Qanon after giving it life}.
\newblock \url{https://nbcnews.to/2FIl0Ky}, 2018.

\bibitem{germanyqanon}
{NBC News}.
\newblock {QAnon supporters join thousands at protest against Germany's
  coronavirus rules}.
\newblock \url{https://nbcnews.to/2HfuIVc}, 2020.

\bibitem{ny2017voat}
{New York Magazine}.
\newblock {The Death of Alt-Right Reddit Is a Good Reminder That the Market
  Does Not Like Nazi Internet}.
\newblock \url{https://nym.ag/2HgNxrc}, 2017.

\bibitem{newell2016user}
E.~Newell, D.~Jurgens, H.~M. Saleem, H.~Vala, J.~Sassine, C.~Armstrong, and
  D.~Ruths.
\newblock User migration in online social networks: A case study on reddit
  during a period of community unrest.
\newblock {\em ICWSM}, 2016.

\bibitem{newshub2017aircrash}
{Newshub}.
\newblock {New theories claim MH370 was `remotely hijacked', buried in
  Antarctica}.
\newblock \url{https://bit.ly/3m7u9f5}, 2017.

\bibitem{papasavva2020raiders}
A.~Papasavva, S.~Zannettou, E.~De~Cristofaro, G.~Stringhini, and J.~Blackburn.
\newblock Raiders of the lost kek: 3.5 years of augmented 4chan posts from the
  politically incorrect board.
\newblock In {\em ICWSM}, 2020.

\bibitem{jigsaw2018perspective}
{Perspective API}.
\newblock \url{https://www.perspectiveapi.com/}, 2018.

\bibitem{popova2019reading}
M.~Popova.
\newblock Reading out of context: pornographic deepfakes, celebrity and
  intimacy.
\newblock {\em Porn Studies}, 2019.

\bibitem{rivers2014ethical}
C.~M. Rivers and B.~L. Lewis.
\newblock {Ethical research standards in a world of big data}.
\newblock {\em F1000Research}, 2014.

\bibitem{See2019FromCT}
{Rose See}.
\newblock {From Crumbs to Conspiracy: Qanon as a community of hermeneutic
  practice}.
\newblock \url{https://scholarship.tricolib.brynmawr.edu/handle/10066/21223},
  2019.

\bibitem{saleem2017web}
H.~M. Saleem, K.~P. Dillon, S.~Benesch, and D.~Ruths.
\newblock {A web of hate: Tackling hateful speech in online social spaces}.
\newblock In {\em TACOS}, 2016.

\bibitem{santos2020mapping}
A.~Santos~Rutschman.
\newblock {Mapping Misinformation in the Coronavirus Outbreak}.
\newblock
  \url{https://www.healthaffairs.org/do/10.1377/hblog20200309.826956/full/},
  2020.

\bibitem{sap2019risk}
M.~Sap, D.~Card, S.~Gabriel, Y.~Choi, and N.~A. Smith.
\newblock The risk of racial bias in hate speech detection.
\newblock In {\em ACL}, 2019.

\bibitem{emerson2020whatisvoat}
{Sarah Emerson}.
\newblock {Months Before Reddit Purge, The\_Donald Users Created a New Home}.
\newblock
  \url{https://onezero.medium.com/months-before-reddit-purge-the-donald-users-created-a-new-home-a732f79e4f04},
  2020.

\bibitem{schabes2020birtherism}
{Schabes, Emily}.
\newblock {Birtherism, Benghazi and QAnon: Why Conspiracy Theories Pose a
  Threat to American Democracy}.
\newblock {\em Student Research}, (158), 2020.

\bibitem{sky2020whatisq}
{Sky News}.
\newblock {What is QAnon? The bizarre pro-Trump conspiracy theory growing ahead
  of the US election}.
\newblock \url{https://bit.ly/3bEqUYL}, 2020.

\bibitem{spacy}
spaCy.
\newblock {Industrial-Strength Natural Language Processing}.
\newblock \url{https://spacy.io/}, 2019.

\bibitem{sternisko2020dark}
A.~Sternisko, A.~Cichocka, and J.~J. Van~Bavel.
\newblock {The Dark Side of Social Movements: Social Identity, Non-conformity,
  and the Lure of Conspiracy Theories}.
\newblock {\em Current opinion in psychology}, 2020.

\bibitem{stuff2018bansq}
{Stuff News}.
\newblock {Reddit bans r/greatawakening board, home to QAnon conspiracy
  theorists}.
\newblock \url{https://bit.ly/31phSei}, 2018.

\bibitem{li2018lda}
{Suzan Li}.
\newblock {Topic Modeling and Latent Dirichlet Allocation (LDA) in Python}.
\newblock
  \url{https://towardsdatascience.com/topic-modeling-and-latent-dirichlet-allocation-in-python-9bf156893c24},
  2018.

\bibitem{swami2012social}
V.~Swami.
\newblock {Social psychological origins of conspiracy theories: The case of the
  Jewish conspiracy theory in Malaysia}.
\newblock {\em Frontiers in Psychology}, 2012.

\bibitem{cancel_voat_contract}
{The Guardian}.
\newblock {Reddit clone Voat dropped by web host for `politically incorrect'
  content}.
\newblock \url{https://bit.ly/34bPgqt}, 2015.

\bibitem{guardian2018qanon}
{The Guardian}.
\newblock {What is QAnon? Explaining the bizarre rightwing conspiracy theory}.
\newblock \url{https://bit.ly/3o8TGX5}, 2018.

\bibitem{QAnoncrimes}
{The Guardian}.
\newblock {QAnon: a timeline of violence linked to the conspiracy theory}.
\newblock \url{https://bit.ly/3oEeXIp}, 2020.

\bibitem{youtube2020qanonban}
{The Guardian}.
\newblock {YouTube announces plans to ban content related to QAnon}.
\newblock \url{https://bit.ly/31jNwK4}, 2020.

\bibitem{capitol}
{The New York Times}.
\newblock {These Are the 5 People Who Died in the Capitol Riot}.
\newblock \url{https://nyti.ms/2XzYXLr}, 2021.

\bibitem{verge2015fph}
{The Verge}.
\newblock {Reddit Bans `Fat People Hate' and Other Subreddits Under new
  Harassment Rules}.
\newblock \url{https://bit.ly/3o6EVnH}, 2015.

\bibitem{verge2015voat}
{The Verge}.
\newblock {Welcome to Voat: Reddit killer, troll haven, and the strange face of
  internet free speech}.
\newblock \url{https://bit.ly/37lRRQJ}, 2015.

\bibitem{verge2018deep}
{The Verge}.
\newblock {Reddit Bans `deepfakes' AI Porn Communities}.
\newblock \url{https://bit.ly/35hS1Wy}, 2018.

\bibitem{wired2018qanon}
{The Verge}.
\newblock {Reddit has banned the QAnon conspiracy subreddit r/GreatAwakening}.
\newblock \url{https://bit.ly/3jfYuXb}, 2018.

\bibitem{verge2020twitterban}
{The Verge}.
\newblock {Getting rid of QAnon won’t be as easy as Twitter might think}.
\newblock \url{https://bit.ly/35dwO0a}, 2020.

\bibitem{paypal}
{The Washington Post}.
\newblock {This is what happens when you create an online community without any
  rules, part 2}.
\newblock \url{https://wapo.st/2ZjsNVH}, 2015.

\bibitem{wp2018banq}
{The Washington Post}.
\newblock {Reddit bans r/greatawakening, the main subreddit for QAnon
  conspiracy theorists}.
\newblock \url{https://wapo.st/2R6Jnnm}, 2018.

\bibitem{thomasid2020}
E.~Thomas and A.~Zhang.
\newblock {ID2020, Bill Gates and the Mark of the Beast: How Covid-19 catalyses
  existing online conspiracy movements}.
\newblock {\em JSTOR}, 2020.

\bibitem{Prooijen2018ThePO}
J.-W. van Prooijen.
\newblock {The psychology of QAnon: Why do seemingly sane people believe
  bizarre conspiracy theories?}
\newblock \url{https://nbcnews.to/379FHcV}, 2018.

\bibitem{vox2019bestseller}
{Vox}.
\newblock {How a conspiracy theory about Democrats drinking children’s blood
  topped Amazon’s best-sellers list}.
\newblock \url{https://bit.ly/31qA8E6}, 2019.

\bibitem{wang2012baselines}
S.~I. Wang and C.~D. Manning.
\newblock Baselines and bigrams: Simple, good sentiment and topic
  classification.
\newblock In {\em ACL}, 2012.

\bibitem{wired2015coon}
{Wired}.
\newblock {Why Reddit banned `CoonTown', and what comes next}.
\newblock
  \url{https://www.wired.co.uk/article/reddit-new-content-policy-update}, 2015.

\bibitem{QAnonbook}
{WWG1WGA}.
\newblock {\em {QAnon: An Invitation to the Great Awakening}}.
\newblock Relentlessly Creative Books, 2018.

\bibitem{zannettou2020quantitative}
S.~Zannettou, J.~Finkelstein, B.~Bradlyn, and J.~Blackburn.
\newblock A quantitative approach to understanding online antisemitism.
\newblock In {\em ICWSM}, 2020.

\bibitem{Zuckerman2019QAnonAT}
E.~Zuckerman.
\newblock Qanon and the emergence of the unreal.
\newblock {\em Journal of Design and Science}, (6), 2019.

\end{thebibliography}

\end{document}